\documentclass[12pt,letterpaper]{article}
\usepackage[a4paper, total={7in, 10in}]{geometry}

\usepackage{graphicx}
\usepackage{helvet}
\usepackage{authblk}
\usepackage{hyperref}
\usepackage{amsmath} 
\usepackage{amssymb} 
\usepackage{orcidlink} 
\usepackage[super,comma,sort&compress]  
   {natbib}
\usepackage[right]{lineno} \linenumbers
\usepackage[title]{appendix}%
\usepackage{float}

\usepackage[utf8]{inputenc} 
\usepackage[T1]{fontenc}    
\usepackage{url}            
\usepackage{booktabs}       
\usepackage{amsfonts}       
\usepackage{nicefrac}       
\usepackage{microtype}      
\usepackage{xcolor}         
\usepackage{multirow}
\usepackage{threeparttable}
\usepackage{amsmath}
\usepackage{subcaption}
\usepackage{makecell}
\usepackage{placeins}
\usepackage[utf8]{inputenc} 
\usepackage{booktabs}      
\usepackage{siunitx}       
\usepackage{caption}       
\usepackage{tabularx}
\usepackage{listings}    
\usepackage{pifont}
\usepackage{array}
\usepackage{makecell}
\usepackage{caption}
\usepackage{pdfpages}
\usepackage{placeins} 
\usepackage{booktabs,multirow,adjustbox,makecell}
\theadset{\bfseries}

\makeatletter
\renewcommand{\maketitle}{\bgroup\setlength{\parindent}{0pt}
\begin{flushleft}
  \textbf{\@title}
  
  \@author
\end{flushleft}\egroup}
\makeatother

\title{Holter-to-Sleep: AI-Enabled Repurposing of Single-Lead ECG for Sleep Phenotyping}
\date{}

\author[1,2\orcidlink{0009-0008-6988-2183}]{Donglin Xie}
\author[3]{Qingshuo Zhao}
\author[4]{Jingyu Wang}
\author[5]{Shijia Geng}
\author[6]{Jiarui Jin}
\author[1,2]{Jun Li}
\author[4]{Rongrong Guo}
\author[1,6]{Guangkun Nie}
\author[1,2]{Gongzheng Tang}
\author[3,7]{Yuxi Zhou}
\author[8,*]{Thomas Penzel}
\author[1,2,9,*,\orcidlink{0000-0001-7521-5127}]{Shenda Hong}

\affil[1]{National Institute of Health Data Science, Peking University, Beijing, China}
\affil[2]{Institute of Medical Technology, Peking University Health Science Center, Beijing, China}
\affil[3]{Department of Computer Science, Tianjin University of Technology, Tianjin, China}
\affil[4]{Department of Respiratory and Critical Care Medicine, Binzhou Medical University Hospital, Binzhou, China}
\affil[5]{Heart Voice Medical Technology, Hefei, Anhui}
\affil[6]{School of Intelligence Science and Technology, Peking University}
\affil[7]{Department of Respiratory, The Second Hospital of Tianjin Medical University, Tianjin, China}
\affil[8]{Interdisciplinary Center of Sleep Medicine, Charité - Universitätsmedizin Berlin, Berlin, Germany}
\affil[9]{Institute for Artificial Intelligence, Peking University, Beijing, China}

\affil[*]{Correspondence: hongshenda@pku.edu.cn, thomas.penzel@charite.de}

\begin{document}

\maketitle



\section*{ABSTRACT}
Sleep disturbances are tightly linked to cardiovascular risk, yet polysomnography (PSG)—the clinical reference standard—remains resource-intensive and poorly suited for multi-night, home-based, and large-scale screening. Single-lead electrocardiography (ECG), already ubiquitous in Holter and patch-based devices, enables comfortable long-term acquisition and encodes sleep-relevant physiology through autonomic modulation and cardiorespiratory coupling. Here, we present a proof-of-concept Holter-to-Sleep framework that, using single-lead ECG as the sole input, jointly supports overnight sleep phenotyping and Holter-grade cardiac phenotyping within the same recording, and further provides an explicit analytic pathway for scalable cardio--sleep association studies. The framework is developed and validated on a pooled multi-center PSG sample of 10,439 studies spanning four public cohorts, with independent external evaluation to assess cross-cohort generalizability, and additional real-world feasibility assessment using overnight patch-ECG recordings via objective--subjective consistency analysis. This integrated design enables robust extraction of clinically meaningful overnight sleep phenotypes under heterogeneous populations and acquisition conditions, and facilitates systematic linkage between ECG-derived sleep metrics and arrhythmia-related Holter phenotypes. Collectively, the Holter-to-Sleep paradigm offers a practical foundation for low-burden, home-deployable, and scalable cardio--sleep monitoring and research beyond traditional PSG-centric workflows.

\section*{KEYWORDS}
Holter-to-Sleep, Single-lead ECG, Scalable home sleep phenotyping, Cardio--sleep association analysis


\section*{Introduction}

Sleep is a fundamental biological process that sustains systemic homeostasis and is essential for cognitive performance, emotional regulation, and metabolic stability. Persistent sleep disturbances are consistently associated with adverse health outcomes, including cardiovascular disease, metabolic dysfunction, and neurocognitive impairment \cite{barone2011diabetes, lal2012neurocognitive, hargens2013association, vandekerckhove2017emotion, vyazovskiy2015sleep, zisapel2007sleep, wang2025phenome}. Accordingly, scalable and accessible sleep monitoring is increasingly important for both clinical screening and population-level sleep health management.

PSG is widely regarded as the clinical gold standard for sleep assessment. By recording multimodal signals---including electroencephalography (EEG), electrooculography (EOG), electromyography (EMG), respiratory airflow, oxygen saturation, and ECG---PSG enables comprehensive characterization of sleep architecture and sleep-related events, such as arousals and disordered breathing\cite{van2011objective, roebuck2013review, mazzotti2018opportunities}. Despite its indispensable role, PSG remains resource-intensive and operationally constrained by specialized equipment, trained personnel, and laborious scoring procedures, limiting its scalability for multi-night monitoring, home deployment, and large-scale screening \cite{de2024state}.

Among candidate modalities for scalable sleep monitoring, single-lead ECG is uniquely positioned to bridge clinical-grade assessment and real-world deployment. Single-lead ECG is already widely available from routine Holter and patch-based wearable devices, supports comfortable long-term recordings at home, and enables continuous, low-burden monitoring across multiple nights. From a physiological perspective, sleep-related autonomic modulation, respiratory load, and arousal responses are reflected in heart rate dynamics and ECG morphology, motivating ECG-based inference of sleep stages and sleep disturbances \cite{sun2020sleep, wei2018research, wei2019multi, li2021transfer, zhao2021dual, utomo2019automatic, tripathi2022ensemble, urtnasan2022deep, lesmana2018sleep, tang2022deep, li2018deep, bozkurt2021sleep, zhu2026artificial}. Importantly, ECG is inherently dual-purpose: if robust sleep characterization can be achieved from the same signal used for cardiac monitoring, a single overnight recording can concurrently provide sleep metrics and Holter-grade cardiac analytics, thereby enabling scalable investigation of cardio--sleep associations.

Prior work has demonstrated the feasibility of ECG-based sleep analysis, spanning conventional feature-based approaches and deep learning models for single-lead ECG sleep staging and sleep event detection \cite{plesinger2018automated, li2020deep, mack2006passive, ragnarsdottir2018automatic, song2015obstructive, bahrami2022sleep, varon2015novel, mendonca2018review, radha2019sleep}. In parallel, the rapid growth of wearable sensing has stimulated renewed interest in home sleep monitoring using ECG and other unobtrusive signals \cite{kwon2021recent, kwon2023home, chen2013unobtrusive}. Nevertheless, substantial gaps remain before single-lead ECG can serve as a comprehensive and generalizable substrate for scalable sleep monitoring, and these gaps directly motivate the present study. First, many ECG-based studies rely on limited-scale datasets with restricted population diversity, which constrains robustness under cross-cohort distribution shifts \cite{kemp2000analysis, ichimaru1999development, goldberger2000physiobank}. Second, sleep is governed by long-range temporal dynamics and structured stage transitions; models that insufficiently capture multi-epoch context may underperform when generalized across centers and cohorts. Third, existing approaches often treat sleep staging, respiratory event detection, and arousal recognition as isolated tasks, preventing shared representation learning and limiting the ability to produce PSG-like holistic outputs. Fourth, external validation and real-world wearable testing remain scarce, leaving uncertainties regarding reliability under practical acquisition conditions. Finally, even when sleep inference is feasible, a systematic pathway to integrate ECG-derived sleep phenotypes with Holter-level arrhythmia analytics for scalable cardio--sleep association studies is largely under-explored.

\begin{figure}
    \centering
    \includegraphics[width=\textwidth]{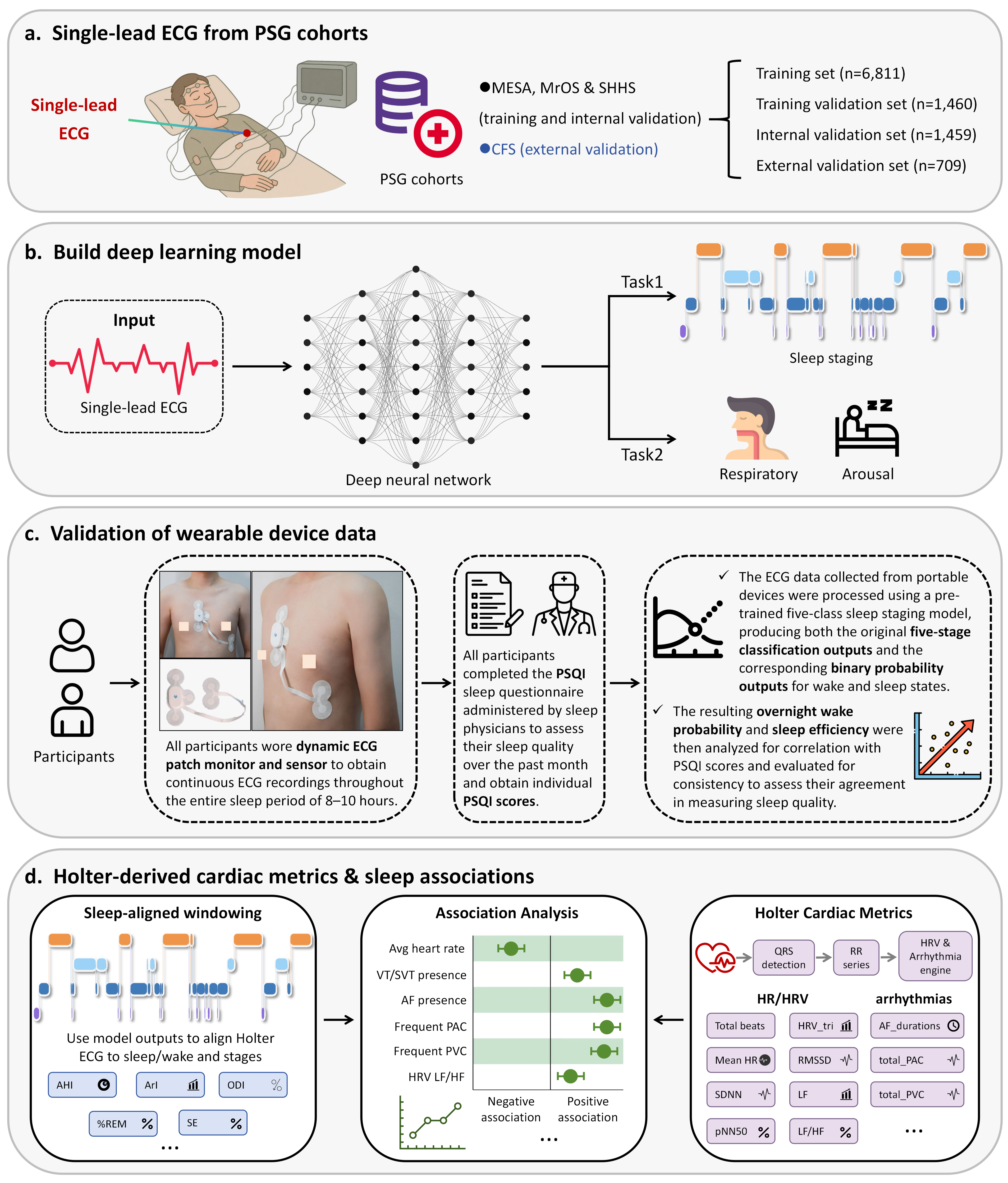}
    \caption{\textbf{Overview of the Holter-to-Sleep study design and analytic workflow.}
    \textbf{a,} Data collection. Single-lead ECG was extracted from multi-center PSG cohorts.
    \textbf{b,} Model development. ECG-based deep learning model for sleep staging and sleep disturbance detection.
    \textbf{c,} Wearable data validation. Participants underwent overnight recordings using a patch-based ECG device, accompanied by PSQI-based subjective sleep quality assessment; model-derived sleep-consistency metrics were used to evaluate real-world agreement.
    \textbf{d,} Holter-derived cardiac metrics and sleep associations. Model outputs were used to align long-duration nocturnal ECG with sleep, enabling joint extraction of sleep metrics and Holter-grade cardiac phenotypes, followed by association analyses to support scalable cardio--sleep phenotyping.}
    \label{fig:overview}
\end{figure}

Here we present a \textbf{Holter-to-Sleep} framework that uses single-lead ECG as the sole input to jointly characterize sleep architecture and clinically relevant sleep disturbances at scale, with rigorous multi-cohort validation and real-world wearable assessment, and further enables Holter-level cardio--sleep association analyses. Figure~\ref{fig:overview} displays an overview of the study, and the key contributions of this study are as follows:
\begin{itemize}
    \item \textbf{Holter-to-Sleep framework:} We establish a Holter-to-Sleep framework demonstrating that single-lead ECG can support holistic sleep characterization for scalable home monitoring.
    \item \textbf{Two-Stage deep learning architecture:} We introduce a Two-Stage deep learning architecture that combines CNN-based morphological representation learning with Transformer-based temporal refinement to better capture long-range sleep dynamics and improve cross-cohort generalization.
    \item \textbf{Unified multi-task learning:} We develop a Unified multi-task learning strategy that simultaneously models sleep staging and detects key sleep disturbances, aligning model outputs with PSG-style reporting.
    \item \textbf{Multi-cohort validation and wearable assessment:} We conduct multi-center training/internal validation and independent external validation on NSRR cohorts \cite{zhang2018national, redline1995familial, chen2015racial, blackwell2011associations, quan1997sleep}, and further perform real-world wearable validation using patch-based ECG with objective--subjective consistency assessment.
    \item \textbf{Holter-level cardio--sleep association analysis:} We perform Holter-level cardio--sleep association analyses by linking ECG-derived sleep phenotypes with arrhythmia-related cardiac phenotypes, providing a scalable analytic pathway for cardio--sleep medicine.
\end{itemize}

\section*{RESULTS}

\subsection*{PSG cohorts}
A total of $10{,}439$ PSG studies were included, and single-lead ECG was extracted from each record as the model input. The combined MESA~\cite{chen2015racial}, MrOS~\cite{blackwell2011associations}, and SHHS~\cite{quan1997sleep} cohorts ($n=9{,}730$) were used for model development and internal evaluation, and were split into a training set ($n=6{,}811$), a training-validation set ($n=1{,}460$), and an internal validation set ($n=1{,}459$). The CFS cohort served as an independent external validation set ($n=709$). Table~\ref{tab:dataset_characteristics} summarizes demographics and PSG-derived sleep and event-burden metrics for each split (continuous variables reported as mean$\pm$s.d.; categorical variables reported as $n$ (\%)). The external cohort differs from the development cohorts in age/sex composition and several sleep-related indices, providing an independent setting for assessing cross-cohort generalizability.

\begin{table}
\centering
\caption{\textbf{Baseline characteristics and PSG-derived sleep metrics across data splits.}
Continuous variables are reported as mean$\pm$s.d.; categorical variables are reported as $n$ (\%). AHI, apnea-hypopnea index; ArI, arousal index; ODI, oxygen desaturation index; TST, total sleep time; TIB, time in bed; WASO, wake after sleep onset.}
\label{tab:dataset_characteristics}

\resizebox{\textwidth}{!}{%
\begin{tabular}{lcccc}
\toprule
\textbf{Metric} & \textbf{Training} ($n=6{,}811$) & \textbf{Train-val} ($n=1{,}460$) & \textbf{Internal val} ($n=1{,}459$) & \textbf{External val} ($n=709$) \\
\midrule
\multicolumn{5}{l}{\textbf{Demographics}}\\
Age (years) & 68.48$\pm$10.89 & 68.38$\pm$11.20 & 68.72$\pm$10.56 & 41.34$\pm$19.45 \\
Female, $n$ (\%) & 2,354 (34.56\%) & 496 (33.97\%) & 495 (33.93\%) & 388 (54.72\%) \\
BMI (kg\,m$^{-2}$) & 27.94$\pm$4.84 & 28.04$\pm$4.87 & 28.02$\pm$4.91 & 32.12$\pm$9.26 \\
\addlinespace

\multicolumn{5}{l}{\textbf{Event burden and oxygen desaturation}}\\
AHI (events\,h$^{-1}$) & 20.21$\pm$17.02 & 20.39$\pm$17.41 & 20.22$\pm$16.98 & 12.30$\pm$16.75 \\
ArI (events\,h$^{-1}$) & 16.49$\pm$6.68 & 16.46$\pm$6.68 & 16.33$\pm$6.24 & 12.75$\pm$6.07 \\
ODI 3\% (events\,h$^{-1}$) & 6.41$\pm$9.74 & 6.62$\pm$9.80 & 6.43$\pm$9.67 & 6.43$\pm$12.50 \\
ODI 4\% (events\,h$^{-1}$) & 3.71$\pm$7.18 & 3.91$\pm$7.34 & 3.74$\pm$7.20 & 4.07$\pm$9.46 \\
\addlinespace

\multicolumn{5}{l}{\textbf{Sleep architecture (percentage of TST)}}\\
N1 / TST (\%) & 7.81$\pm$6.71 & 7.88$\pm$6.62 & 8.05$\pm$6.89 & 5.14$\pm$4.94 \\
N2 / TST (\%) & 58.50$\pm$11.28 & 58.91$\pm$11.18 & 58.42$\pm$11.47 & 56.08$\pm$13.07 \\
N3 / TST (\%) & 14.45$\pm$11.24 & 13.94$\pm$11.14 & 14.18$\pm$11.24 & 21.20$\pm$13.77 \\
NREM / TST (\%) & 80.76$\pm$6.59 & 80.73$\pm$6.61 & 80.65$\pm$6.53 & 82.42$\pm$7.27 \\
REM / TST (\%) & 19.24$\pm$6.59 & 19.27$\pm$6.61 & 19.35$\pm$6.53 & 17.58$\pm$7.27 \\
\addlinespace

\multicolumn{5}{l}{\textbf{Sleep continuity and fragmentation}}\\
TIB (min) & 528.69$\pm$84.83 & 527.74$\pm$84.38 & 527.70$\pm$84.87 & 535.92$\pm$49.52 \\
TST (min) & 352.73$\pm$67.33 & 351.86$\pm$67.44 & 353.77$\pm$65.08 & 353.34$\pm$70.88 \\
Sleep efficiency (\%) & 68.02$\pm$15.14 & 67.93$\pm$15.09 & 68.29$\pm$14.62 & 66.24$\pm$13.32 \\
Wake / TIB (\%) & 31.98$\pm$15.14 & 32.07$\pm$15.09 & 31.71$\pm$14.62 & 33.76$\pm$13.32 \\
Sleep latency (min) & 53.31$\pm$57.80 & 55.74$\pm$58.21 & 53.41$\pm$59.39 & 98.67$\pm$68.77 \\
WASO (min) & 84.76$\pm$61.99 & 83.24$\pm$62.61 & 83.48$\pm$60.82 & 76.47$\pm$60.79 \\
\bottomrule
\end{tabular}%
} 
\end{table}

\subsection*{Consistent sleep staging and epoch-level sleep event detection performance with strong cross-cohort generalization}

\begin{figure}
    \centering
    \includegraphics[width=0.98\textwidth]{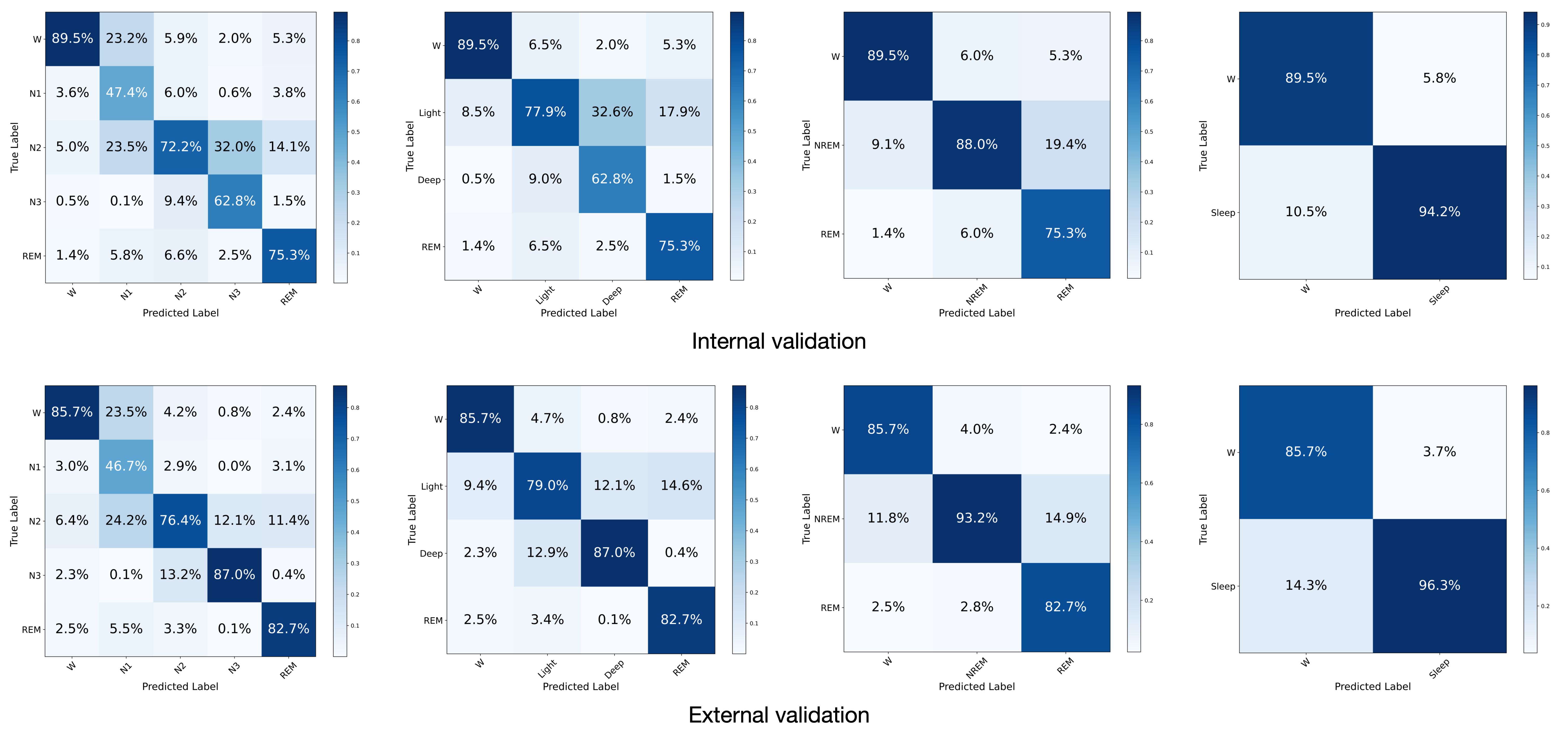}
    \caption{\textbf{Cross-cohort generalization of single-lead ECG sleep staging on internal and external validation.}
    The top row shows internal validation and the bottom row shows external validation; from left to right are confusion matrices for 5-class, 4-class, 3-class, and 2-class staging. Each cell reports the epoch count, with the percentage in parentheses normalized by the predicted class. For 4-class staging, Light merges N1+N2 and Deep corresponds to N3; for 3-class staging, NREM merges N1--N3; for 2-class staging, Sleep merges N1--N3 and REM.}
    \label{fig:staging_generalization}
\end{figure}

\begin{figure}
    \centering
    \includegraphics[width=0.98\textwidth]{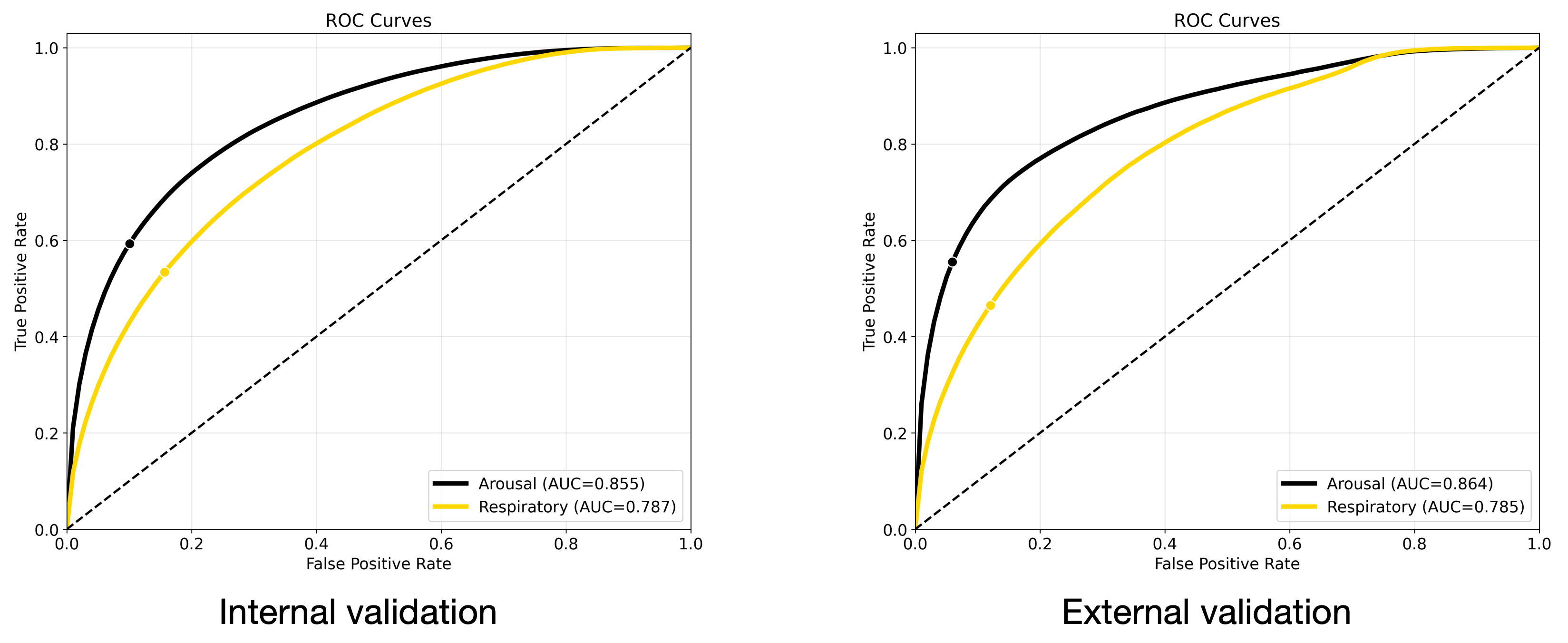}
    \caption{\textbf{Cross-cohort generalization of sleep event detection.}
    ROC curves are shown for arousal and respiratory event detection; the left panel corresponds to internal validation and the right panel to external validation. The area under the ROC curve (AUC) is reported in each panel.}
    \label{fig:event_generalization}
\end{figure}

To assess cross-cohort generalizability, we evaluated the proposed model on the held-out internal validation set from MESA/MrOS/SHHS and the independent external validation cohort (CFS). Confusion matrices for four staging granularities are shown in Figure~\ref{fig:staging_generalization}, and the corresponding quantitative metrics with 95\% confidence intervals (CIs) are summarized in Table~\ref{tab:staging_metrics}. In 4-class staging (W/Light/Deep/REM), the model achieved $\kappa=0.701$ [0.700, 0.702] on internal validation and $\kappa=0.744$ [0.743, 0.745] on external validation (Macro AUC: 0.944 [0.944, 0.944] and 0.959 [0.959, 0.959], respectively). For sleep disturbance detection, ROC curves are shown in Figure~\ref{fig:event_generalization} and performance metrics with 95\% CIs are reported in Table~\ref{tab:event_metrics}. Arousal detection achieved an AUC of 0.855 [0.855, 0.856] on internal validation and 0.864 [0.863, 0.865] on external validation, while respiratory event detection achieved an AUC of 0.787 [0.786, 0.788] and 0.785 [0.784, 0.787], respectively.

To provide qualitative context for single-lead ECG sleep staging, we curated representative ECG-based sleep-staging studies and summarized their reported datasets, PSG sample sizes, and headline metrics in Appendix~\ref{app:comparison}. Because cohorts, label sets, and evaluation protocols vary across studies, this summary is intended for context rather than head-to-head benchmarking; likewise, we did not compile cross-study comparisons for sleep-event detection given substantial heterogeneity in event definitions, time scales, and disorder taxonomies.

\begin{table}
\centering
\caption{\textbf{Cross-cohort performance for ECG-based sleep staging at multiple granularities.}
Values are reported as point estimates with 95\% CIs in brackets.
For 4-class staging, Light merges N1+N2 and Deep corresponds to N3; for 3-class staging, NREM merges N1--N3; for 2-class staging, Sleep merges N1--N3 and REM.}
\label{tab:staging_metrics}
\small
\setlength{\tabcolsep}{6pt}
\renewcommand{\arraystretch}{1.15}
\begin{tabular}{lcccc}
\toprule
\textbf{Class setting} & \textbf{Accuracy $\uparrow$} & \textbf{Weighted F1 $\uparrow$} & \textbf{Kappa $\uparrow$} & \textbf{Macro AUC $\uparrow$} \\
\midrule
\multicolumn{5}{l}{\textbf{Internal validation}} \\
5-class & 0.773 [0.772, 0.773] & 0.761 [0.760, 0.762] & 0.670 [0.668, 0.670] & 0.930 [0.930, 0.930] \\
4-class & 0.804 [0.803, 0.804] & 0.801 [0.801, 0.802] & 0.701 [0.700, 0.702] & 0.944 [0.944, 0.944] \\
3-class & 0.870 [0.869, 0.870] & 0.869 [0.869, 0.870] & 0.776 [0.775, 0.777] & 0.962 [0.962, 0.962] \\
2-class & 0.926 [0.926, 0.926] & 0.926 [0.926, 0.926] & 0.834 [0.834, 0.835] & 0.974 [0.974, 0.975] \\
\addlinespace
\multicolumn{5}{l}{\textbf{External validation (CFS)}} \\
5-class & 0.812 [0.811, 0.813] & 0.801 [0.800, 0.802] & 0.729 [0.728, 0.731] & 0.946 [0.945, 0.946] \\
4-class & 0.826 [0.825, 0.827] & 0.822 [0.821, 0.822] & 0.744 [0.743, 0.745] & 0.959 [0.959, 0.959] \\
3-class & 0.892 [0.891, 0.893] & 0.892 [0.891, 0.893] & 0.814 [0.813, 0.815] & 0.975 [0.974, 0.975] \\
2-class & 0.924 [0.923, 0.924] & 0.925 [0.924, 0.925] & 0.833 [0.832, 0.834] & 0.978 [0.977, 0.978] \\
\bottomrule
\end{tabular}
\end{table}

\begin{table}
\centering
\caption{\textbf{Cross-cohort performance for ECG-based sleep event detection.}
Values are reported as point estimates with 95\% CIs in brackets. AUC, area under the receiver operating characteristic curve.}
\label{tab:event_metrics}
\small
\setlength{\tabcolsep}{8pt}
\renewcommand{\arraystretch}{1.15}
\begin{tabular}{lcc}
\toprule
\textbf{Metric} & \textbf{Arousal} & \textbf{Respiratory} \\
\midrule
\multicolumn{3}{l}{\textbf{Internal validation}} \\
Accuracy $\uparrow$ & 0.852 [0.852, 0.853] & 0.798 [0.798, 0.799] \\
Precision $\uparrow$ & 0.518 [0.517, 0.520] & 0.368 [0.366, 0.369] \\
Recall $\uparrow$ & 0.593 [0.591, 0.595] & 0.534 [0.532, 0.536] \\
Specificity $\uparrow$ & 0.899 [0.899, 0.900] & 0.843 [0.843, 0.844] \\
F1-score $\uparrow$ & 0.553 [0.552, 0.555] & 0.435 [0.434, 0.437] \\
AUC $\uparrow$ & 0.855 [0.855, 0.856] & 0.787 [0.786, 0.788] \\
\addlinespace
\multicolumn{3}{l}{\textbf{External validation (CFS)}} \\
Accuracy $\uparrow$ & 0.895 [0.894, 0.895] & 0.838 [0.837, 0.839] \\
Precision $\uparrow$ & 0.558 [0.554, 0.561] & 0.301 [0.298, 0.304] \\
Recall $\uparrow$ & 0.555 [0.552, 0.558] & 0.465 [0.461, 0.469] \\
Specificity $\uparrow$ & 0.940 [0.940, 0.941] & 0.880 [0.879, 0.880] \\
F1-score $\uparrow$ & 0.556 [0.553, 0.559] & 0.365 [0.363, 0.368] \\
AUC $\uparrow$ & 0.864 [0.863, 0.865] & 0.785 [0.784, 0.787] \\
\bottomrule
\end{tabular}
\end{table}

\subsection*{ECG emerges as a promising modality for scalable sleep staging after channel-wise comparison}

To contextualize the role of single-lead ECG in sleep staging and quantify the benefit of the proposed two-stage model, CNN with Transformer-based temporal refinement, we performed a channel-wise comparison on the CFS cohort (Table~\ref{tab:channel_comparison}). Adding the Transformer consistently improved staging performance across all channels relative to the CNN-only (Net1D) baseline. Notably, for ECG-2, $\kappa$ increased from 0.621 to 0.748 (a 20.45\% relative increase), with Accuracy$=0.823$, Weighted F1$=0.819$, and Macro AUC$=0.949$. Under single-channel settings, ECG-2 approached the performance of channels more directly associated with sleep staging, while the multimodal reference set (C4+LOC+EMG1) achieved the highest overall performance.

\begin{table}
\centering
\caption{\textbf{Channel-wise comparison on the CFS cohort highlights ECG as a competitive modality for scalable sleep staging.}
Performance is reported for 5-class sleep staging using individual PSG channels.
Net1D denotes the CNN-only baseline, and CNN+Trans denotes the proposed two-stage model with temporal refinement.
Relative $\kappa$ increase is computed as $(\kappa_{\mathrm{CNN+Trans}}-\kappa_{\mathrm{CNN}})/\kappa_{\mathrm{CNN}}\times 100\%$.}
\label{tab:channel_comparison}
\small
\setlength{\tabcolsep}{6pt}
\renewcommand{\arraystretch}{1.15}

\resizebox{\textwidth}{!}{%
\begin{tabular}{llccccc}
\toprule
\textbf{Channel} & \textbf{Method} & \textbf{Accuracy $\uparrow$} & \textbf{Weighted F1 $\uparrow$} & \textbf{$\kappa$ $\uparrow$} & \textbf{Macro AUC $\uparrow$} & \textbf{Relative $\kappa$ increase} \\
\midrule
\multirow{2}{*}{C4+LOC+EMG1} & Net1D & 0.843 & 0.842 & 0.782 & 0.962 & -- \\
 & Net1D+Trans & 0.886 & 0.881 & 0.838 & 0.977 & +7.16\% \\
\addlinespace

\multirow{2}{*}{EEG-C4} & Net1D & 0.827 & 0.822 & 0.759 & 0.955 & -- \\
 & Net1D+Trans & 0.873 & 0.869 & 0.820 & 0.974 & +8.04\% \\
\addlinespace

\multirow{2}{*}{EOG-L} & Net1D & 0.832 & 0.829 & 0.767 & 0.956 & -- \\
 & Net1D+Trans & 0.878 & 0.875 & 0.826 & 0.975 & +7.69\% \\
\addlinespace

\multirow{2}{*}{EMG1} & Net1D & 0.705 & 0.690 & 0.580 & 0.880 & -- \\
 & Net1D+Trans & 0.787 & 0.778 & 0.700 & 0.936 & +20.69\% \\
\addlinespace

\multirow{2}{*}{ECG-2} & Net1D & 0.725 & 0.718 & 0.621 & 0.903 & -- \\
 & Net1D+Trans & 0.823 & 0.819 & 0.748 & 0.949 & +20.45\% \\
\addlinespace

\multirow{2}{*}{Airflow} & Net1D & 0.482 & 0.459 & 0.247 & 0.694 & -- \\
 & Net1D+Trans & 0.632 & 0.610 & 0.454 & 0.836 & +83.81\% \\
\bottomrule
\end{tabular}%
}
\end{table}

\subsection*{Real-world wearable data validation: ECG-derived sleep consistency metrics correlate with PSQI}

\begin{figure}
    \centering
    \includegraphics[width=0.98\textwidth]{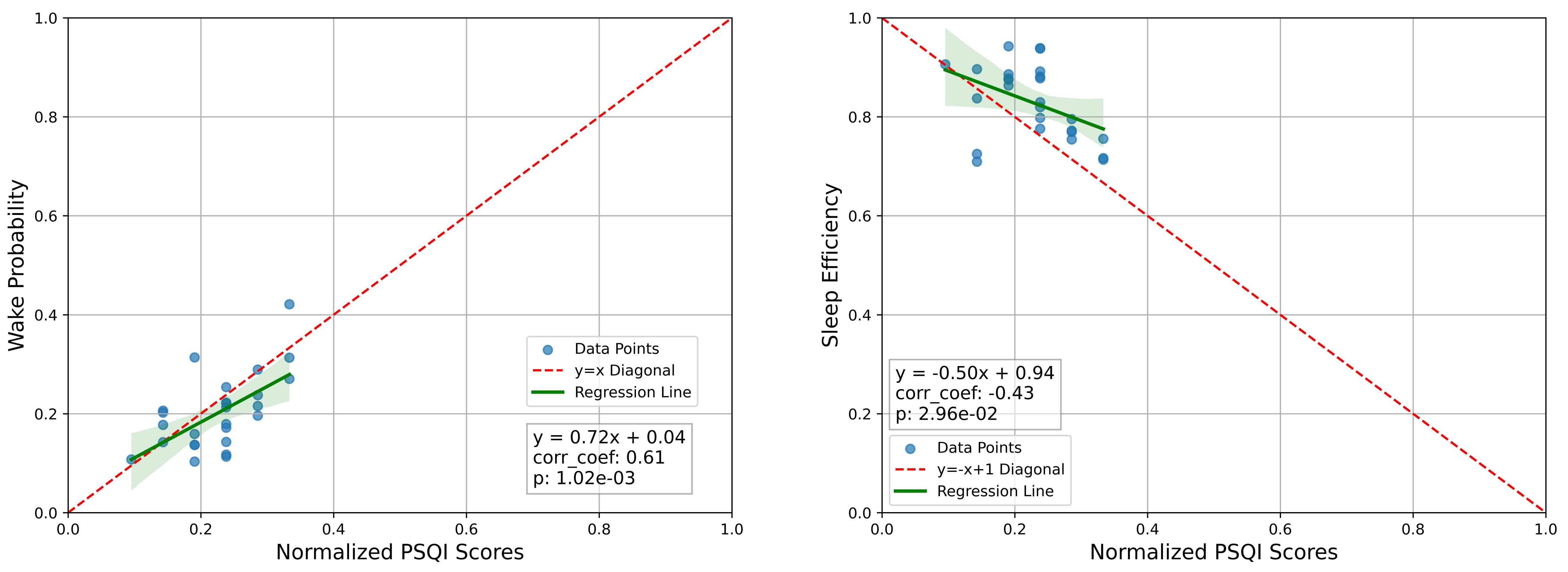}
    \caption{\textbf{Real-world wearable validation of ECG-derived sleep consistency metrics against PSQI.}
    Scatter plots show associations between normalized PSQI scores and model-derived overnight metrics.
    \textbf{Left:} wake probability, defined as the overnight mean of epoch-wise predicted wake probabilities.
    \textbf{Right:} sleep efficiency estimated from model-inferred sleep--wake information.
    Solid lines indicate linear regression with 95\% confidence bands. Pearson's correlation coefficient ($r$) and two-sided $p$ values are shown in each panel.}
    \label{fig:wearable_psqi}
\end{figure}

In the patch-based wearable ECG cohort, epoch-level PSG labels were unavailable. We therefore evaluated real-world consistency by comparing model-derived overnight metrics with subjective sleep quality assessed by the Pittsburgh Sleep Quality Index (PSQI) \cite{buysse1989pittsburgh}(Fig.~\ref{fig:wearable_psqi}). Specifically, wake probability was defined as the overnight mean of epoch-wise wake probabilities, and sleep efficiency was estimated from model-inferred sleep--wake information. In $n=26$ participants, wake probability was positively correlated with normalized PSQI ($r=0.61$, $p=1.02\times 10^{-3}$), whereas sleep efficiency was negatively correlated with normalized PSQI ($r=-0.43$, $p=2.96\times 10^{-2}$).

\subsection*{ECG-derived sleep phenotypes reveal cross-cohort heart--sleep associations}

\begin{figure}
    \centering
    \includegraphics[width=0.98\textwidth, height=0.88\textheight, keepaspectratio]{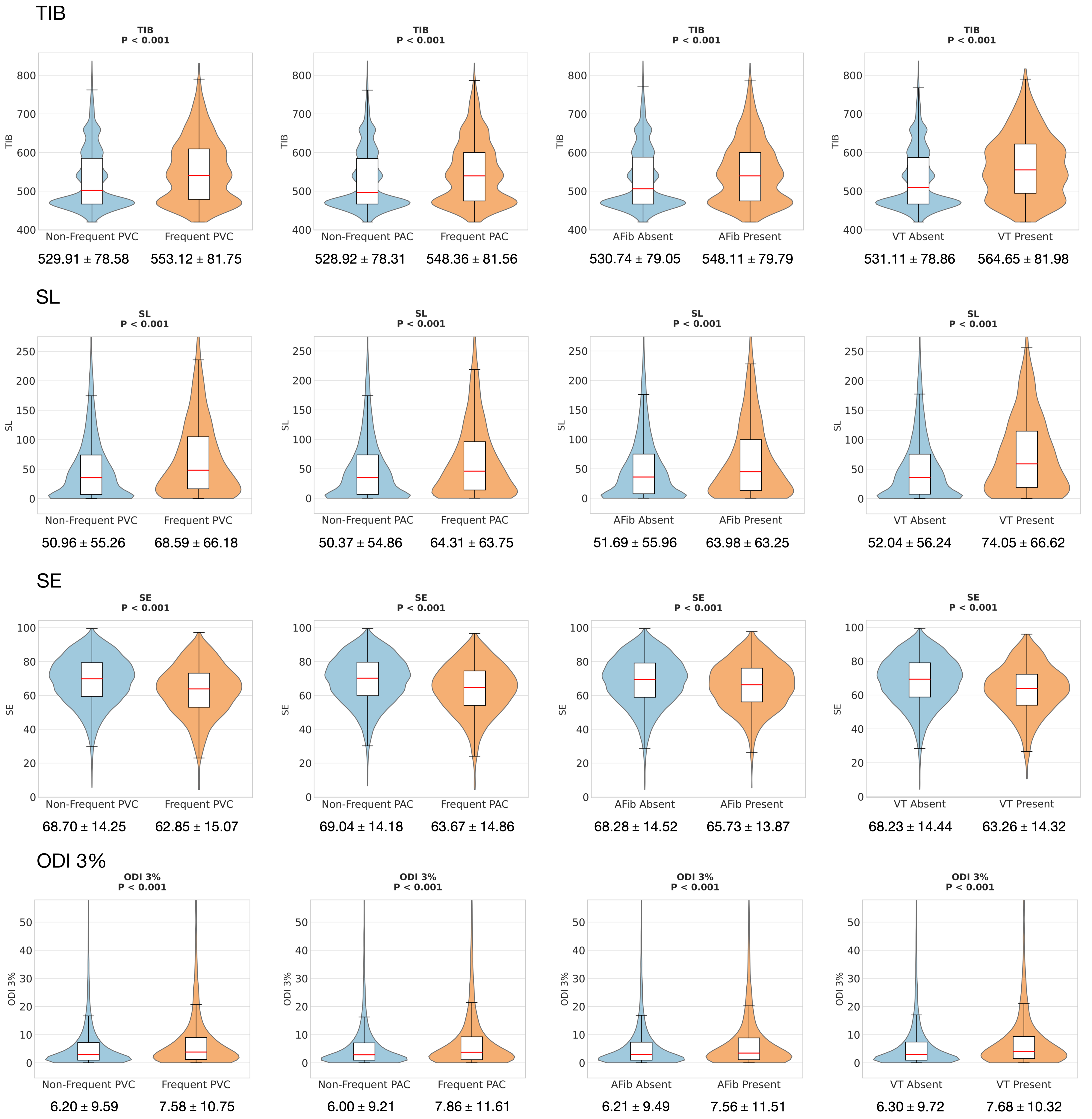}
    \caption{\textbf{ECG-derived sleep phenotypes differ across Holter arrhythmia-related strata.}
    Violin plots show distributions of time in bed (TIB), sleep latency (SL), sleep efficiency (SE), and oxygen desaturation index at 3\% (ODI 3\%).
    Participants are stratified by (i) frequent premature ventricular contractions (PVC burden $\geq 21$ events\,h$^{-1}$) versus non-frequent PVC,
    (ii) frequent premature atrial contractions (PAC burden $\geq 30$ events\,h$^{-1}$) versus non-frequent PAC,
    (iii) atrial fibrillation (AF) presence during sleep versus absence, and (iv) ventricular tachycardia (VT) presence during sleep versus absence.
    Embedded boxplots indicate the median and interquartile range. Values below each panel denote mean$\pm$s.d.
    $P$ values from between-group comparisons are annotated in each panel. TIB and SL are reported in minutes, SE in \%, and ODI 3\% in events\,h$^{-1}$.}
    \label{fig:holter_associations}
\end{figure}

\begingroup
\setlength{\emergencystretch}{2em} 
\sloppy

Across the pooled multi-cohort PSG sample, we derived overnight sleep phenotypes from single-lead ECG model inference and compared their distributions across arrhythmia-related strata (Fig.\ \ref{fig:holter_associations}). Relative to their corresponding reference groups, individuals with arrhythmia phenotypes or higher ectopy burden consistently exhibited poorer sleep, characterized by longer TIB, longer SL, lower SE, and higher ODI 3\%; all stratified comparisons annotated in the figure were statistically significant ($P<0.001$).

For frequent premature ventricular contraction (PVC), TIB increased from 529.91$\pm$78.58 to 553.12$\pm$81.75\,min, SL increased from 50.96$\pm$55.26 to 68.59$\pm$66.18\,min, SE decreased from 68.70$\pm$14.25\% to 62.85$\pm$15.07\%, and ODI 3\% increased from 6.20$\pm$9.59 to 7.58$\pm$10.75.
For frequent premature atrial contraction (PAC), TIB increased from 528.92$\pm$78.31 to 548.36$\pm$81.56\,min, SL increased from 50.37$\pm$54.86 to 64.31$\pm$63.75\,min, SE decreased from 69.04$\pm$14.18\% to 63.67$\pm$14.86\%, and ODI 3\% increased from 6.00$\pm$9.21 to 7.86$\pm$11.61.
For atrial fibrillation (AF) during sleep, TIB increased from 530.74$\pm$79.05 to 548.11$\pm$79.79\,min, SL increased from 51.69$\pm$55.96 to 63.98$\pm$63.25\,min, SE decreased from 68.28$\pm$14.52\% to 65.73$\pm$13.87\%, and ODI 3\% increased from 6.21$\pm$9.49 to 7.56$\pm$11.51.
For ventricular tachycardia (VT), the differences were more pronounced: TIB increased from 531.11$\pm$78.86 to 564.65$\pm$81.98\,min, SL increased from 52.04$\pm$56.24 to 74.05$\pm$66.62\,min, SE decreased from 68.23$\pm$14.44\% to 63.26$\pm$14.32\%, and ODI 3\% increased from 6.30$\pm$9.72 to 7.68$\pm$10.32.

Collectively, these results indicate that ECG-inferred sleep phenotypes can robustly capture discriminative sleep differences aligned with Holter-grade arrhythmia phenotypes in a cross-cohort setting. Further subgroup analyses of phenotypes and complete results can be found in the Appendix~\ref{app:joint_tables}.

\endgroup

\subsection*{From algorithm to application: an integrated web dashboard for ECG-based sleep \& Holter cardiac reporting}

\begin{figure}
    \centering
    \includegraphics[width=0.98\textwidth, height=0.85\textheight, keepaspectratio]{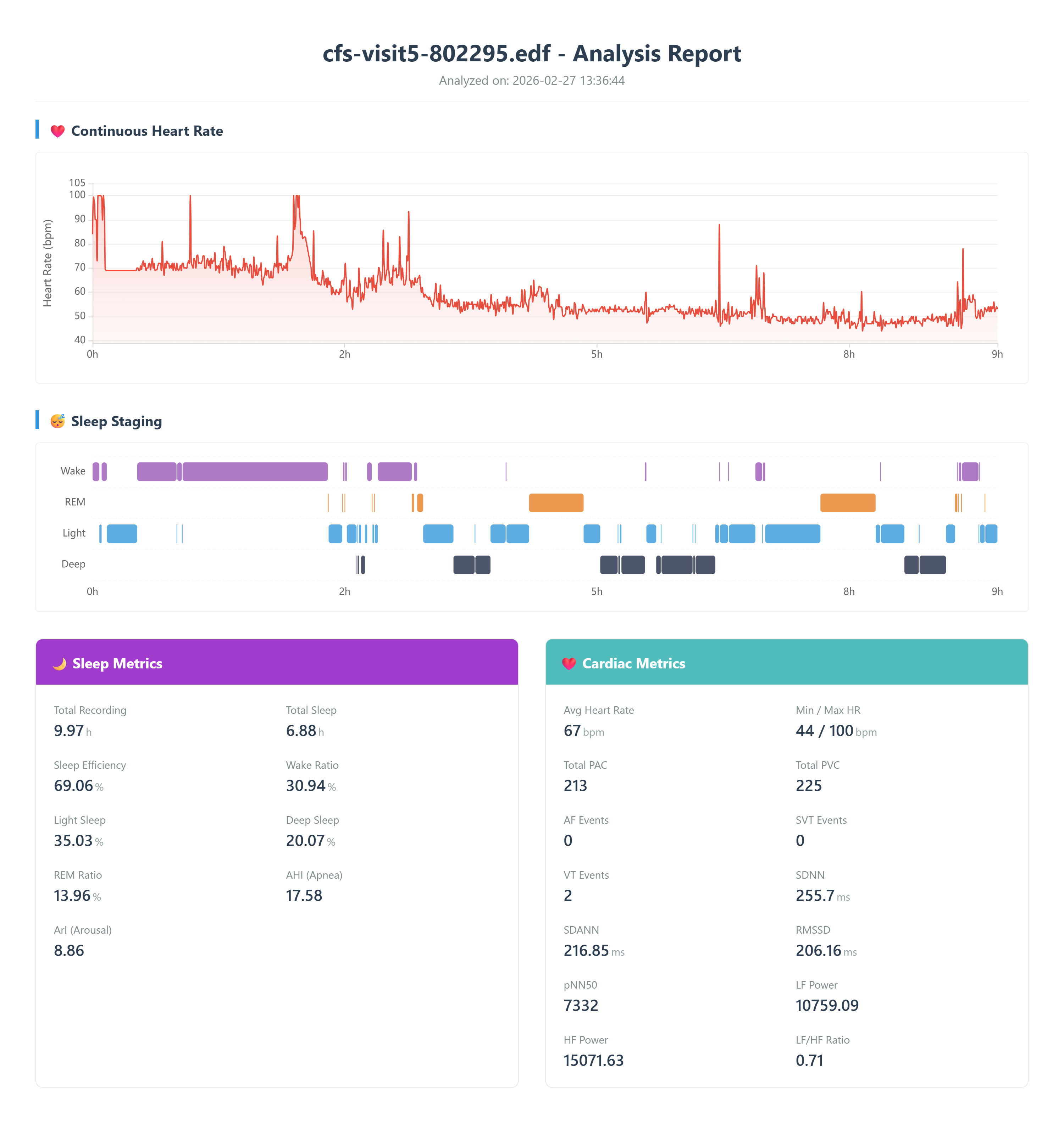}
    \caption{\textbf{An integrated web dashboard for ECG-based sleep and Holter cardiac reporting.}
    The dashboard supports raw EDF input, automatically extracts the ECG channel, and integrates the proposed sleep inference pipeline with Holter-grade cardiac analytics. The interface provides (top) an overnight heart-rate trend, (middle) a sleep-stage timeline, and (bottom) summary sleep metrics and cardiac metrics (including HRV and arrhythmia-related summaries) in a consolidated report.}
    \label{fig:web_dashboard}
\end{figure}

To facilitate practical use and result interpretation, we implemented an integrated web dashboard that combines ECG-based sleep inference with Holter-grade long-term ECG analytics (Fig.~\ref{fig:web_dashboard}). The system accepts raw EDF recordings, automatically parses the ECG channel, and generates a consolidated overnight report including continuous heart-rate trends, a sleep-stage timeline, and summary sleep and cardiac metrics, enabling unified review of nocturnal cardio--sleep phenotypes.

\section*{DISCUSSION}

This study proposes and validates a proof-of-concept \emph{Holter-to-Sleep} framework that enables simultaneous sleep phenotyping and Holter-grade cardiac phenotyping using single-lead ECG alone, together with an explicit pathway for downstream cardio--sleep association analyses. In contrast to prior work that typically targets a single sleep task or relies on multi-channel PSG signals, our framework provides a scalable paradigm that integrates nocturnal sleep assessment and cardiac risk characterization within the same data source and workflow. This unified design is particularly relevant to home-based, lightweight, and low-burden monitoring scenarios, and it lays a methodological and engineering foundation for scalable cardio--sleep research, longitudinal health management, and potential screening or follow-up applications.

Regarding methodological validity, we systematically evaluated ECG-based sleep staging, arousal detection, and respiratory event detection on multi-center cohorts totaling more than 10,000 PSG studies, and further assessed generalizability via internal validation and an independent external cohort. Although the study is not intended as a head-to-head benchmarking against existing methods, the scale of the data and the consistent cross-cohort performance support that the proposed approach is extensible and transferable across heterogeneous populations, providing a strong basis for deployment in larger cohorts and more complex real-world settings. Algorithmically, the proposed design aims to maximally exploit sleep-relevant physiological information embedded in single-lead ECG. First, we fused time-domain, frequency-domain, and time--frequency representations to enhance sensitivity to autonomic modulation, rhythmic dynamics, and non-stationary components. Second, a CNN backbone captures local morphology and short-term dynamics, while a Transformer module models longer-range temporal dependencies, improving discrimination of sleep architecture and event boundaries. Third, multi-epoch concatenation incorporates contextual information from neighboring epochs and leverages temporal consistency to stabilize epoch-level predictions. Ablation studies in Appendix~\ref{app:ablation} indicate that each component contributes to the overall performance.

The channel-wise comparison (Table~\ref{tab:channel_comparison}) provides complementary evidence for single-lead ECG as a strong and pragmatic modality for scalable sleep staging from the perspective of sensor availability, information content, and deployability. Under the stringent single-channel setting, ECG-2 with temporal refinement achieved performance close to channels more directly associated with conventional staging (e.g., EEG and EOG), while clearly outperforming respiratory surrogates that are more vulnerable to real-world signal degradation (e.g., airflow). Notably, the relative gain brought by the Transformer was particularly pronounced for ECG, suggesting that stage-related cardiac cues are expressed less as isolated within-epoch patterns and more as temporally structured dynamics---reflecting the minute-scale unfolding of sleep continuity, transition regularities, and autonomic/respiratory modulation of heart-rate fluctuations. This observation motivates the two-stage design adopted in this work: learning robust epoch-level representations while explicitly modeling contextual dependencies and sequence-level constraints using attention. From a translational standpoint, single-lead ECG is routinely available in PSG and constitutes a standard sensing modality in Holter and patch-based monitors; thus, leveraging ECG enables competitive sleep staging without the additional acquisition burden of dedicated neurophysiological electrodes, supporting longitudinal follow-up and population-scale cardio--sleep phenotyping. We emphasize that multimodal PSG remains the reference standard for maximal diagnostic fidelity; the present work targets cost-effective, scalable screening and phenotyping rather than replacing full clinical polysomnography.

Beyond PSG-based validation, we further examined real-world feasibility using a dynamic ECG patch monitor (HeartVoice, Anhui, China) to acquire overnight recordings in an unconstrained setting. Because epoch-level PSG labels are unavailable in this wearable cohort, we derived interpretable overnight metrics from model outputs (overnight wake probability and sleep efficiency) and evaluated their consistency with subjective sleep quality assessed by the PSQI. The observed associations suggest that the proposed approach can provide informative sleep-related summaries under real-world conditions without PSG supervision, supporting its potential for broader deployment and longitudinal studies.

From a clinical and public health perspective, jointly assessing sleep and cardiac phenotypes is necessary. In the Holter cohort, arrhythmia phenotypes and higher ectopy burden were associated with a consistent pattern of worse sleep, including increased time in bed, prolonged sleep latency, reduced sleep efficiency, and elevated hypoxemia burden. These findings suggest that conventional Holter analyses alone may not fully capture nocturnal physiology, and that incorporating sleep phenotypes within the same monitoring framework can help reveal clinically meaningful interactions and risk signals. A single-lead ECG solution that simultaneously supports sleep and cardiac phenotyping is therefore well aligned with the need for low-burden, home-based, continuous cardio--sleep monitoring.

Several additional strengths are noteworthy. We leveraged multi-center cohorts and performed external validation, supporting reproducibility and scalability. The end-to-end pipeline spans data ingestion, sleep inference, cardiac analytics, association analysis, and reporting, lowering the engineering barrier for translation. Moreover, the implemented web dashboard enables automated parsing of raw EDF recordings and generation of integrated overnight reports, facilitating practical adoption by researchers and clinicians and accelerating translation from proof-of-concept to real-world workflows.

This study also has limitations. First, the wearable validation relied on PSQI as a reference. PSQI is subjective and reflects sleep quality over approximately one month, whereas our ECG recordings covered a single night; thus, the wearable analysis primarily characterizes correlations between single-night model-derived metrics and recent subjective sleep quality rather than strict epoch-level accuracy in real-world sleep staging. Nevertheless, this approach provides a pragmatic starting point for feasibility assessment in the absence of PSG labels. Future work should incorporate multi-night recordings, sleep diaries, and/or additional wearable modalities to strengthen ecological validity. Second, supplementary experiments with cohort-specific training (Appendix~\ref{appendixA}) show reduced performance when training and evaluating on MESA or SHHS alone. This may reflect the higher prevalence of cardiovascular comorbidities in these cohorts, leading to more complex and heterogeneous ECG morphologies and rhythms that challenge learning of stable sleep-relevant representations within a single cohort. Future directions include comorbidity-aware stratification, domain adaptation, and explicit robustness mechanisms for arrhythmias and noise to improve performance in high-risk populations.

In summary, by using single-lead ECG as the sole input, we establish an extensible Holter-to-Sleep framework that supports multi-task sleep inference, integrates Holter-grade cardiac phenotyping, and enables cardio--sleep association analyses, providing a feasible pathway toward low-burden, home-based continuous cardio--sleep monitoring.

\section*{METHODS}


\subsection*{Datasets and data preprocessing}

\subsubsection*{Datasets and data annotation}
\begin{figure}
    \centering
    \includegraphics[width=0.98\textwidth, height=0.85\textheight, keepaspectratio]{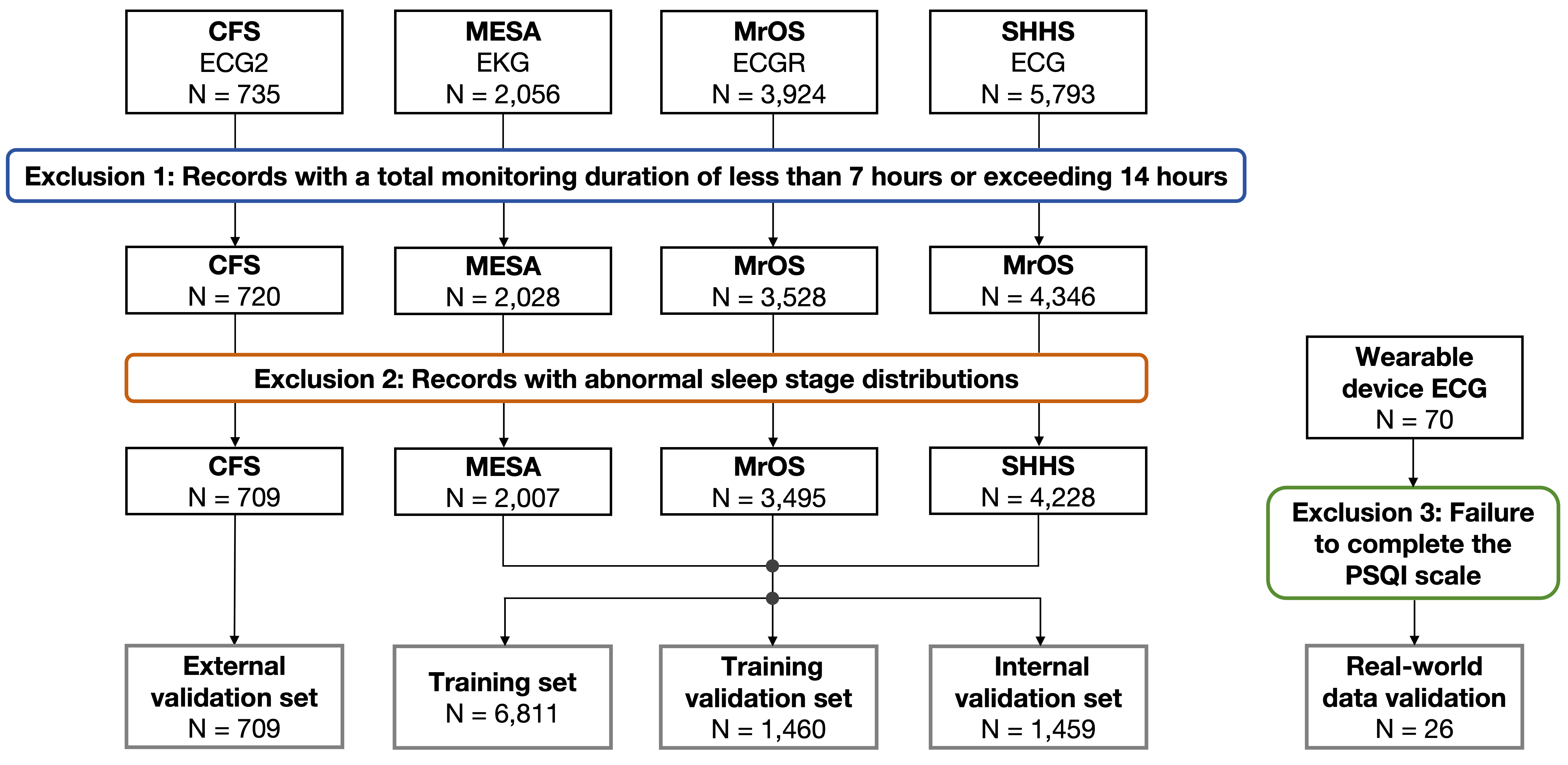}
    \caption{\textbf{Study dataset construction and screening flow.}
    Public PSG cohorts included CFS (ECG2, $N=735$), MESA (EKG, $N=2056$), SHHS (ECG, $N=5793$), and MrOS (ECGR, $N=3924$).
    \textbf{Exclusion 1:} recordings with total monitoring duration $<7$\,h or $>14$\,h; remaining samples were CFS $N=720$, MESA $N=2028$, SHHS $N=4346$, and MrOS $N=3528$.
    \textbf{Exclusion 2:} recordings with abnormal sleep-stage distributions; remaining samples were CFS $N=709$, MESA $N=2007$, SHHS $N=4228$, and MrOS $N=3495$.
    CFS was used as an independent external validation set ($N=709$).
    MESA, SHHS, and MrOS were pooled and split into a training set ($N=6811$), a training-validation set ($N=1460$), and an internal validation set ($N=1459$).
    For real-world wearable ECG, $N=70$ recordings were collected; \textbf{Exclusion 3:} failure to complete the PSQI questionnaire; the final real-world validation subset comprised $N=26$.}
    \label{fig:dataset_flow}
\end{figure}

In our study, we assembled a unified dataset for model development, validation, and real-world evaluation by integrating single-lead ECG signals from four large public PSG cohorts and by collecting additional long-term wearable ECG recordings. The public cohorts included MESA, MrOS, SHHS, and CFS. MESA, MrOS, and SHHS were used for model training and internal validation, whereas CFS served as an independent external validation cohort. Real-world wearable data were acquired using a patch-based dynamic ECG monitor, yielding 70 overnight recordings; after questionnaire completion and quality control, these data were used for real-world consistency evaluation. To ensure data quality and cross-cohort comparability, we applied unified inclusion/exclusion criteria with quality control on monitoring duration and sleep-stage distributions. The complete screening process and cohort-specific sample sizes are summarized in Fig.~\ref{fig:dataset_flow}.

We strictly aligned single-lead ECG signals extracted from PSG with the corresponding sleep annotations by parsing XML annotation files matched to the EDF recordings. Three 30-s epoch--level label sequences were generated for each record: sleep staging labels, arousal event labels, and respiratory event labels. Sleep stages followed standard 30-s scoring with Wake, N1, N2, N3, and REM; for consistency, N3 and N4 annotations were merged into N3. Arousal labels were defined as a binary sequence, where an epoch was labeled as 1 if any arousal event occurred within the 30-s window and 0 otherwise. Respiratory labels were also defined in a binary manner: an epoch was labeled as 1 if hypopnea or any apnea event (obstructive, central, or mixed) occurred within the 30-s window, and 0 otherwise. To ensure complete temporal coverage and accurate signal--label alignment, we computed the total number of 30-s epochs for each PSG recording and constructed label sequences of equal length accordingly. Because the beginning and end of recordings are often affected by device setup/termination and signal instability, we excluded the first and last 60 epochs (approximately 60 minutes in total; 30 minutes at each end) from each ECG record to mitigate boundary noise. Ultimately, each recording yielded a time-aligned signal matrix together with sleep stage, arousal, and respiratory event labels for downstream modeling and analysis.

\subsubsection*{Data preprocessing}
A standardized preprocessing pipeline was applied to single-lead ECG signals extracted from PSG recordings to improve signal quality and cross-dataset consistency, including resampling, filtering, baseline correction, and normalization. All ECG signals were first resampled to 100\,Hz to remove sampling-rate variability across cohorts. A 0.5--40\,Hz bandpass filter was then used to suppress low-frequency drift and high-frequency noise, preserving the principal ECG components. Baseline drift was subsequently corrected by fitting and removing a first-order polynomial, thereby reducing slow baseline fluctuations associated with electrode contact variation or motion. To enhance comparability across recordings, robust Z-score normalization was performed  to attenuate the influence of outliers. Finally, amplitude normalization linearly scaled each standardized signal to the range $[-1,1]$, mitigating inter-record amplitude discrepancies and providing uniform inputs for downstream model training.

\subsubsection*{Construction of frequency and time--frequency features}
The neural network input comprises two components: the time-domain ECG waveform and a set of frequency and time--frequency features. The time-domain waveform corresponds to a 30-s segment. For each 30-s segment, the Power Spectral Density (PSD) was first estimated using the Welch method. By segmenting the signal, applying windowing, and performing the Fourier transform, Welch’s approach reduces the influence of noise on spectral estimation and yields frequency components with their corresponding spectral power. Based on the estimated spectrum, nine physiologically relevant frequency-domain features were extracted. Specifically, band powers in Delta (0.5--4\,Hz), Theta (4--8\,Hz), and Alpha (8--13\,Hz) ranges were computed to characterize energy distribution across bands, and total spectral power was calculated to reflect the overall energy level. The peak frequency was obtained as the frequency with the maximum spectral power. In addition, the spectral centroid (power-weighted mean frequency) and spectral bandwidth (standard deviation around the centroid) were computed to summarize the center and spread of the spectral distribution. Spectral shape was further quantified using spectral flatness, where higher flatness indicates a more noise-like spectrum, and spectral entropy was computed on the normalized spectrum to quantify randomness and complexity.

Time--frequency features were additionally derived using wavelet transforms to capture localized signal characteristics across multiple time and frequency scales. Wavelet analysis provides multi-resolution representations and is well suited for non-stationary physiological signals. In this study, a Daubechies wavelet (db4) was adopted as the basis function, and a four-level Discrete Wavelet Transform (DWT) was applied to each 30-s ECG segment, yielding five sets of coefficients: approximation coefficients at level 5 and detail coefficients at levels 1--4. For each coefficient level, energy (sum of squared coefficients) was computed to quantify scale-specific energy contributions. Furthermore, statistical descriptors, including mean, standard deviation, variance, maximum, and minimum, were extracted to characterize amplitude variation and local morphology across scales. Finally, the nine frequency-domain features and the 30 wavelet-derived time--frequency features were concatenated to form a 39-dimensional handcrafted feature vector; the complete feature list is provided in Appendix~\ref{app:handcrafted_features}.

\subsection*{Development of deep learning model}
\begin{figure}
    \centering
    \includegraphics[width=0.98\textwidth, height=0.85\textheight, keepaspectratio]{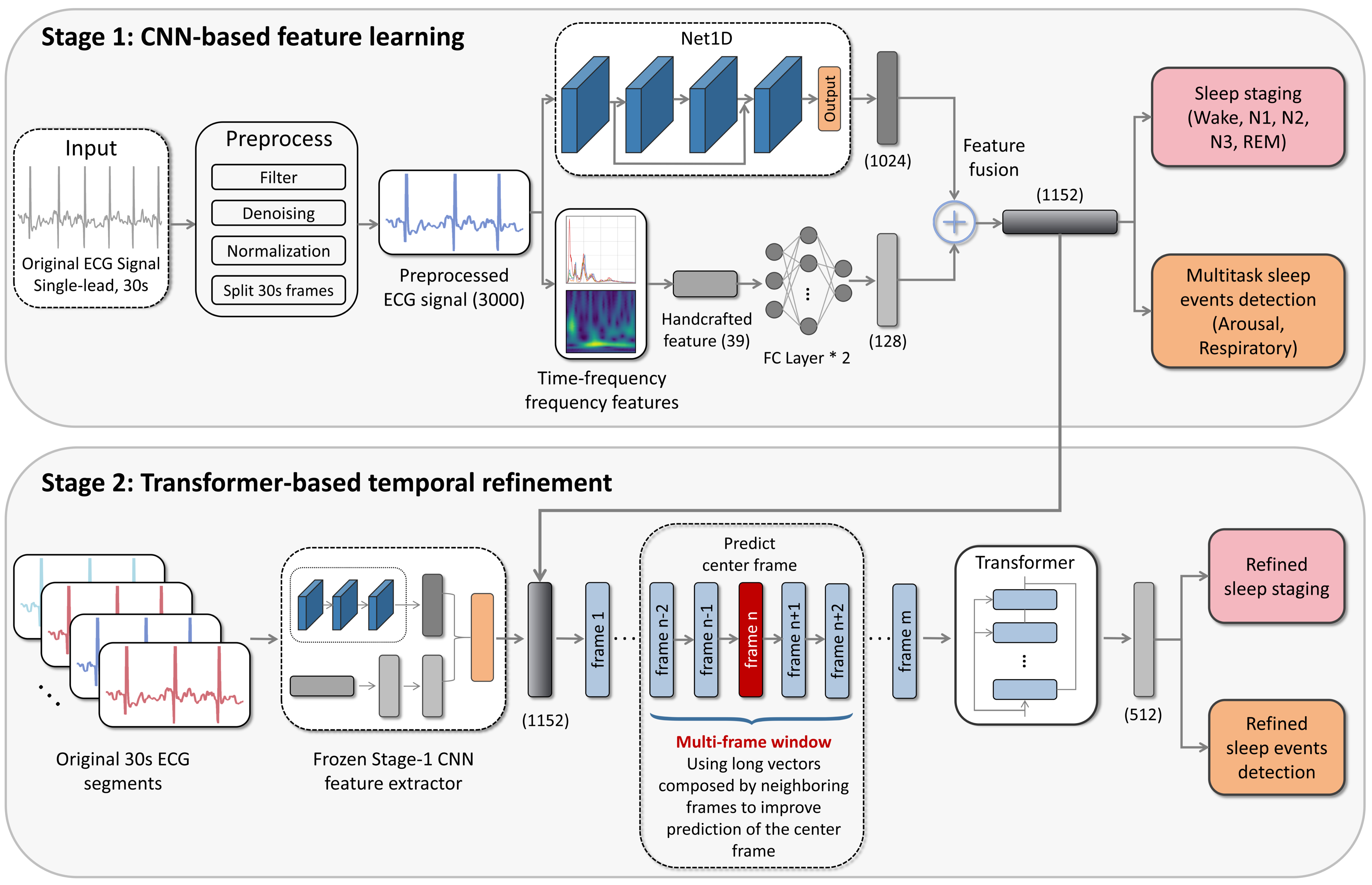}
    \caption{\textbf{Two-stage single-lead ECG framework for sleep staging and event detection.}
    Stage~1 takes 30-s single-lead ECG epochs as input. After preprocessing, a 1D CNN (Net1D) learns deep representations. Frequency/time--frequency features are also extracted and embedded, and then fused with CNN features to produce a shared representation for five-class sleep staging and multi-task sleep event detection, yielding initial predictions.
    Stage~2 constructs a multi-epoch window from neighboring epochs and feeds the sequence of Stage~1 features into a Transformer. The Stage~1 feature extractor is frozen, and the Transformer performs temporal refinement by predicting the center epoch, producing refined sleep staging and event detection outputs.}
    \label{fig:model_architecture}
\end{figure}

We propose a two-stage deep learning framework that takes single-lead ECG as input and jointly performs sleep staging and sleep event detection (Fig.~\ref{fig:model_architecture}). The overall design follows a representation learning--temporal refinement paradigm. Stage~1 employs a CNN to learn robust morphology- and rhythm-related representations from each 30-s epoch, and fuses them with frequency/time--frequency descriptors to form a unified feature vector. Stage~2 freezes the Stage~1 feature extractor and uses a Transformer to model multi-epoch context, improving temporal consistency and boundary discrimination for the center epoch.


\textbf{Stage 1: CNN representation learning with multi-domain feature fusion.} 
For each preprocessed 30-s single-lead ECG epoch, the time-domain signal is fed into a 1D CNN (Net1D) backbone---adapted from the architecture proposed by Hong \emph{et al.}~\cite{hong2020holmes}---to capture morphological and rhythmic patterns, yielding a high-dimensional deep representation. Concurrently, the corresponding frequency-domain features are projected into a compatible embedding space. The deep representation and the handcrafted spectral/time--frequency embeddings are then fused at the feature level to obtain a shared representation for downstream predictions.

\textbf{Task formulation: sleep staging and multi-task event detection.}
Using the shared representation, sleep staging is formulated as a five-class classification task. For sleep events, we adopt a shared-weight multi-task learning head to simultaneously detect arousals and respiratory-related events, leveraging potential physiological relatedness across tasks and improving training efficiency. Outputs from Stage~1 can be directly used as initial predictions.


\textbf{Stage 2: Transformer-based temporal refinement.} 
To exploit the temporal continuity inherent in sleep architecture, we construct a sliding window of neighboring 30-s epochs. The feature sequence extracted from Stage~1 is fed into a Transformer encoder to predict the label of the center epoch~\cite{vaswani2017attention}. During Stage~2, the Stage~1 CNN feature extractor is frozen, and only the Transformer and its prediction heads are optimized. This paradigm enables the model to retain robust local representations while significantly enhancing long-range dependency modeling and stage-transition awareness. This temporal refinement strategy is applied to both sleep staging and sleep-event detection. Specifically, the sleep staging task leverages the sequential context of the epoch-level embeddings, while event detection is formulated as a multitask learning problem utilizing a shared Transformer encoder with parallel classification heads for the binary tasks. Both tasks employ a fixed-length local context window and perform classification based on the representation of the center token. Detailed model architectures and hyperparameters are provided in Appendix~\ref{app:ablation}. The refined outputs from this stage serve as the final predictions for evaluation, derived-metric computation, and subsequent downstream analyses.

\textbf{Training objective and loss functions.}
No specialized loss engineering was introduced beyond standard classification objectives. Sleep staging is optimized using the categorical cross-entropy loss:
\begin{equation}
L_{\text{sleep}} = - \sum_{i=1}^{5} y_i \log(\hat{y_i})
\end{equation}
where $y_i$ and $\hat{y_i}$ denote the one-hot ground-truth label and the predicted probability for class $i$, respectively. For sleep event detection, a multi-task optimization strategy is adopted, where the overall loss is the sum of the arousal loss and the respiratory-event loss:
\begin{equation}
L_{\text{multi-task}} = L_{\text{arousal}} + L_{\text{respiratory}} 
\end{equation}

\subsection*{Data splitting, thresholding, and uncertainty estimation}
Train/validation/test splits were created at the recording level, such that all epochs from the same overnight recording were confined to a single split; the default split ratio was 70\%/15\%/15\% with a fixed random seed for reproducibility. For multiclass sleep staging, discrete labels were obtained by argmax over softmax probabilities. For binary event detection, decision thresholds were determined on the validation set via a grid search (100 equally spaced candidates in $t\in[0.01,0.99]$) by maximizing the F1-score and were then fixed for test evaluation. Metric uncertainty was quantified using nonparametric bootstrap; the resampling unit was the epoch (30-s segment), with 1000 bootstrap replicates and 95\% confidence intervals.

\subsection*{Evaluation metrics}
Model performance was evaluated using task-appropriate classification metrics for sleep staging and sleep event detection. For sleep staging, we report overall Accuracy, Weighted F1-score, Cohen’s Kappa (\(\kappa\)), and macro-averaged ROC AUC (Macro AUC). Kappa is used as a primary staging metric to quantify agreement between predictions and reference annotations while accounting for class imbalance. For sleep event detection (binary classification for arousal and respiratory events), we report Accuracy, Precision, Recall, Specificity, F1-score, and ROC AUC, where AUC serves as the primary indicator of discriminative performance. To quantify the stability and uncertainty of the reported results, 95\% confidence intervals (95\% CI) for key metrics were computed using bootstrapping.

\subsection*{Real-world validation of sleep consistency using wearable ECG data}

To validate the feasibility of the proposed algorithm in real-world sleep monitoring, we conducted a consistency analysis experiment using wearable single-lead ECG devices. The wake probability throughout the night was used as a surrogate metric for sleep quality, and the relationship between this model-derived metric and self-reported PSQI scores was assessed to evaluate the consistency between algorithmic predictions and subjective sleep quality perceptions. A total of 70 healthy participants were recruited for this study. Each participant wore a dynamic single-lead ECG patch for continuous overnight ECG recording. The following morning, participants were asked to complete the PSQI, which assesses subjective sleep quality on a scale from 0 to 21, with higher scores indicating poorer sleep quality. Complete PSQI questionnaires were obtained from 26 participants, who were included in the PSQI-related analyses.

Since single-lead ECG does not provide manual PSG-equivalent annotations, the sleep monitoring algorithm first classified 30-second ECG segments into five sleep stages. To compute the probability of wakefulness across the entire night, we extracted the predicted wake probability from the softmax output of the five-class model for each segment, regardless of its final class label. The per-segment wake probability is denoted as:

\begin{equation}
P_{\text{Wake}, i} = \text{softmax}_{\text{Wake}}(x_i)
\end{equation}

where \(x_i\) represents the input to the model for the \(i\)-th 30-second epoch, and \(P_{\text{Wake}, i}\) denotes the wake probability of the \(i\)-th 30-second segment. The average wake probability over the entire night was computed as:

\begin{equation}
P_{\text{Wake}} = \frac{1}{N} \sum_{i=1}^{N} P_{\text{Wake},i}
\end{equation}

where \(N\) represents the total number of 30-second windows during the night. To align the wake probability range with the PSQI scoring scale, the PSQI scores were normalized by dividing by 21. This normalized PSQI score was then used for consistency analysis.

Additionally, we introduced SE as an additional metric, calculated by summing the total duration of non-wake states over the entire night and dividing by the total monitoring duration. Both the normalized PSQI score and the model-derived metrics were then analyzed for consistency.

To evaluate the relationship between the two model-derived metrics—wake probability and SE—and PSQI scores, the Pearson correlation coefficient (\(r\)) and p-value were computed to quantify their linear association. Furthermore, a least squares regression model was fitted for each metric, and a 95\% confidence interval was calculated to assess the robustness of the regression analysis:

\begin{equation}
y = a \times \mathit{PSQI}_{\text{normalized}} + b
\end{equation}

where \(y\) denotes the model-derived metric, \(a\) is the regression slope, and \(b\) is the intercept.

The full PSQI questionnaire can be found in Appendix~\ref{app:psqi}.

\subsection*{Joint analysis of Holter-level cardiac metrics and sleep metrics}

To assess the associations between ECG-derived sleep phenotypes and Holter-level cardiac phenotypes, we constructed two synchronized feature sets for each PSG recording: a sleep-metric set and a cardiac-metric set. The sleep-metric set comprised 16 variables capturing sleep-disordered breathing burden, arousal burden, and sleep architecture/efficiency. The cardiac-metric set comprised 32 Holter-level variables including heart-rate summary statistics, supraventricular/ventricular ectopy burden and rhythm-event summaries, as well as time- and frequency-domain heart-rate variability (HRV) features. Holter-level metric computation followed established practice and was implemented by adapting prior validated pipelines, with specific reference to the CardioLearn framework by Hong \emph{et al.}~\cite{hong2020cardiolearn} and the implementation described by Fu \emph{et al.}~\cite{fu2021artificial}. Full metric definitions are provided in Appendix~\ref{app:metric_list}.

We then performed joint analysis using a cardiac-phenotype--stratified comparison strategy, in which sleep phenotypes were compared across groups defined by cardiac phenotypes. Two grouping schemes were used depending on the cardiac variable type: (1) binary/event-based stratification---for rhythm events or indicator variables (e.g., atrial fibrillation (AF) occurrence during sleep and ventricular tachycardia (VT) occurrence), participants were stratified into event-present vs.\ event-absent groups; and (2) threshold-based burden stratification---for quantitative burden metrics (e.g., premature atrial contractions (PACs) and premature ventricular contractions (PVCs)), event counts were normalized to an hourly rate (events/h), and participants were stratified into ``frequent'' vs.\ ``non-frequent'' groups using predefined thresholds. Consistent with the Results, frequent PVC was defined as a PVC burden $\geq 21$~events/h and frequent PAC as a PAC burden $\geq 30$~events/h.

For each stratification, we conducted between-group statistical comparisons of sleep metrics to identify sleep features most strongly associated with cardiac phenotypes. Two-sided tests were applied for continuous sleep metrics, and significance levels were reported. Distributions were visualized using box/violin plots. This workflow enables systematic screening of sleep-phenotype differences anchored to Holter-level cardiac phenotypes within a unified PSG cohort, thereby characterizing clinically meaningful cardio--sleep associations.

\subsection*{Web dashboard for integrated sleep--cardiac reporting}

To facilitate practical usability assessment and result visualization, we integrated the proposed ECG-based sleep analysis model with the Holter-level cardiac metric computation module into a user-facing web dashboard. The system allows users to upload EDF (European Data Format) files and select the target single-lead ECG channel. It then automatically performs signal ingestion and quality control, segmentation and preprocessing, model inference, and the computation and aggregation of sleep and cardiac metrics, ultimately generating an overnight integrated sleep--cardiac report. The report includes an overnight heart-rate trend, a sleep-stage timeline, key sleep and arousal/respiratory metrics. The implementation of the dashboard and the end-to-end pipeline are provided in the public repository: \url{https://github.com/xdl1216/Holter-to-Sleep}.

\newpage

\section*{RESOURCE AVAILABILITY}
\subsection*{Data availability}
The public sleep cohorts analyzed during the current study are available in the National Sleep Research Resource (NSRR) repository at \url{https://sleepdata.org/}.

\subsection*{Code availability}
The code for the Holter-to-Sleep framework and the integrated web dashboard are available at \url{https://github.com/xdl1216/Holter-to-Sleep} and \url{https://github.com/PKUDigitalHealth/Holter-to-Sleep}.

\section*{ACKNOWLEDGMENTS}

This work was supported by the National Natural Science Foundation of China (62102008, 62172018), CCF-Tencent Rhino-Bird Open Research Fund (CCF-Tencent RAGR20250108), CCF-Zhipu Large Model Innovation Fund (CCF-Zhipu202414), PKU-OPPO Fund (BO202301, BO202503), Research Project of Peking University in the State Key Laboratory of Vascular Homeostasis and Remodeling (2025-SKLVHR-YCTS-02), Beijing Municipal Science and Technology Commission (Z251100000725008). The authors thank all collaborators and participating institutions for their support and contributions to this research.

\section*{AUTHOR CONTRIBUTIONS}
Donglin Xie conceived and designed the study and implemented the algorithms and software. Qingshuo Zhao integrated, curated, and preprocessed the datasets. Jingyu Wang planned the real-world ECG data collection protocol and provided clinical guidance. Shijia Geng and Rongrong Guo prepared the ECG acquisition devices, performed the real-world ECG recordings, and collected PSQI questionnaires. Jiarui Jin, Jun Li, Guangkun Nie and Gongzheng Tang assisted with result interpretation and manuscript revision. Yuxi Zhou contributed to methodological discussions and manuscript editing. Thomas Penzel and Shenda Hong supervised the project, provided overall technical guidance, and critically revised the manuscript. All authors reviewed and approved the final manuscript.

\section*{DECLARATION OF INTERESTS}


The authors declare no competing interests.

\newpage


\bibliography{reference}

\bigskip


\newpage
\begin{appendices}

\onecolumn

\section{Cohort-specific training and validation results}\label{appendixA}

\vspace{-0.5em}

\noindent\begin{minipage}{\linewidth}
    \centering
    \captionof{table}{Sleep staging performance when training and validating the model independently within each cohort. (Point estimates with 95\% CI).}
    \label{tab:appendixA_staging}
    \scriptsize
    \setlength{\tabcolsep}{5pt}
    \renewcommand{\arraystretch}{1.25}
    \resizebox{\textwidth}{!}{%
    \begin{tabular}{llcccc}
    \toprule
    \textbf{Cohort} & \textbf{Class} & \textbf{Accuracy} $\uparrow$ & \textbf{Weighted F1} $\uparrow$ & $\boldsymbol{\kappa}$ $\uparrow$ & \textbf{Macro AUC} $\uparrow$ \\
    \midrule
    \multirow{4}{*}{CFS} & 5-class & 0.841 [0.839, 0.843] & 0.835 [0.834, 0.837] & 0.776 [0.774, 0.779] & 0.959 [0.959, 0.960] \\
     & 4-class & 0.856 [0.854, 0.857] & 0.856 [0.854, 0.858] & 0.792 [0.789, 0.794] & 0.971 [0.971, 0.972] \\
     & 3-class & 0.910 [0.909, 0.912] & 0.911 [0.909, 0.912] & 0.845 [0.843, 0.848] & 0.981 [0.980, 0.981] \\
     & 2-class & 0.945 [0.944, 0.946] & 0.945 [0.944, 0.946] & 0.875 [0.872, 0.877] & 0.984 [0.984, 0.985] \\
    \midrule
    \multirow{4}{*}{MESA} & 5-class & 0.648 [0.647, 0.650] & 0.624 [0.623, 0.626] & 0.478 [0.477, 0.480] & 0.849 [0.848, 0.849] \\
     & 4-class & 0.699 [0.698, 0.701] & 0.686 [0.685, 0.688] & 0.518 [0.516, 0.520] & 0.864 [0.864, 0.865] \\
     & 3-class & 0.758 [0.757, 0.760] & 0.755 [0.754, 0.757] & 0.584 [0.582, 0.587] & 0.881 [0.881, 0.882] \\
     & 2-class & 0.847 [0.846, 0.848] & 0.846 [0.845, 0.847] & 0.672 [0.670, 0.674] & 0.910 [0.909, 0.911] \\
    \midrule
    \multirow{4}{*}{MrOS} & 5-class & 0.839 [0.838, 0.840] & 0.827 [0.826, 0.828] & 0.758 [0.757, 0.760] & 0.954 [0.954, 0.954] \\
     & 4-class & 0.863 [0.862, 0.864] & 0.863 [0.862, 0.864] & 0.788 [0.786, 0.789] & 0.970 [0.970, 0.971] \\
     & 3-class & 0.900 [0.899, 0.900] & 0.900 [0.899, 0.901] & 0.831 [0.830, 0.832] & 0.977 [0.977, 0.977] \\
     & 2-class & 0.929 [0.928, 0.929] & 0.929 [0.928, 0.929] & 0.853 [0.852, 0.854] & 0.980 [0.980, 0.980] \\
    \midrule
    \multirow{4}{*}{SHHS} & 5-class & 0.646 [0.645, 0.647] & 0.628 [0.626, 0.629] & 0.467 [0.465, 0.469] & 0.848 [0.848, 0.849] \\
     & 4-class & 0.672 [0.671, 0.673] & 0.663 [0.662, 0.664] & 0.486 [0.484, 0.488] & 0.862 [0.861, 0.862] \\
     & 3-class & 0.784 [0.783, 0.785] & 0.780 [0.779, 0.781] & 0.582 [0.579, 0.584] & 0.891 [0.890, 0.892] \\
     & 2-class & 0.897 [0.896, 0.898] & 0.893 [0.892, 0.893] & 0.680 [0.678, 0.682] & 0.920 [0.919, 0.921] \\
    \bottomrule
    \end{tabular}
    }
\end{minipage}

\vspace{1em} 

\noindent\begin{minipage}{\linewidth}
    \centering
    \captionof{table}{\textbf{Arousal Detection Performance:} Independent training and validation within each cohort (Point estimates with 95\% CI).}
    \label{tab:appendixA_arousal}
    \scriptsize
    \setlength{\tabcolsep}{4pt}
    \renewcommand{\arraystretch}{1.2}
    \resizebox{\textwidth}{!}{%
    \begin{tabular}{lcccccc}
    \toprule
    \textbf{Dataset} & \textbf{Accuracy} $\uparrow$ & \textbf{Precision} $\uparrow$ & \textbf{Recall} $\uparrow$ & \textbf{Specificity} $\uparrow$ & \textbf{F1-Score} $\uparrow$ & \textbf{AUC} $\uparrow$ \\
    \midrule
    CFS & 0.911 [0.910, 0.912] & 0.618 [0.611, 0.625] & 0.628 [0.621, 0.636] & 0.949 [0.947, 0.950] & 0.623 [0.618, 0.629] & 0.914 [0.912, 0.916] \\
    MESA & 0.812 [0.811, 0.813] & 0.438 [0.435, 0.441] & 0.568 [0.564, 0.571] & 0.859 [0.858, 0.860] & 0.495 [0.492, 0.498] & 0.808 [0.806, 0.809] \\
    MrOS & 0.891 [0.890, 0.891] & 0.574 [0.571, 0.577] & 0.617 [0.615, 0.620] & 0.932 [0.931, 0.932] & 0.595 [0.592, 0.597] & 0.901 [0.901, 0.902] \\
    SHHS & 0.843 [0.842, 0.844] & 0.551 [0.549, 0.554] & 0.622 [0.620, 0.625] & 0.891 [0.890, 0.891] & 0.585 [0.583, 0.587] & 0.854 [0.853, 0.856] \\
    \bottomrule
    \end{tabular}
    }
\end{minipage}

\vspace{1em}

\noindent\begin{minipage}{\linewidth}
    \centering
    \captionof{table}{\textbf{Respiratory Event Detection Performance:} Independent training and validation within each cohort (Point estimates with 95\% CI).}
    \label{tab:appendixA_respiratory}
    \scriptsize
    \setlength{\tabcolsep}{4pt}
    \renewcommand{\arraystretch}{1.2}
    \resizebox{\textwidth}{!}{%
    \begin{tabular}{lcccccc}
    \toprule
    \textbf{Dataset} & \textbf{Accuracy} $\uparrow$ & \textbf{Precision} $\uparrow$ & \textbf{Recall} $\uparrow$ & \textbf{Specificity} $\uparrow$ & \textbf{F1-Score} $\uparrow$ & \textbf{AUC} $\uparrow$ \\
    \midrule
    CFS & 0.868 [0.867, 0.870] & 0.325 [0.319, 0.332] & 0.490 [0.482, 0.499] & 0.904 [0.902, 0.906] & 0.391 [0.384, 0.398] & 0.841 [0.838, 0.844] \\
    MESA & 0.747 [0.746, 0.748] & 0.173 [0.171, 0.176] & 0.447 [0.442, 0.451] & 0.778 [0.777, 0.779] & 0.250 [0.247, 0.253] & 0.686 [0.683, 0.688] \\
    MrOS & 0.861 [0.860, 0.862] & 0.237 [0.234, 0.239] & 0.390 [0.386, 0.393] & 0.899 [0.898, 0.900] & 0.295 [0.292, 0.297] & 0.793 [0.791, 0.794] \\
    SHHS & 0.717 [0.715, 0.718] & 0.468 [0.466, 0.470] & 0.657 [0.655, 0.659] & 0.738 [0.736, 0.739] & 0.547 [0.545, 0.548] & 0.769 [0.768, 0.770] \\
    \bottomrule
    \end{tabular}
    }
\end{minipage}

\vspace{1em}

\noindent\begin{minipage}{\linewidth}
    \centering
    \includegraphics[width=0.9\textwidth]{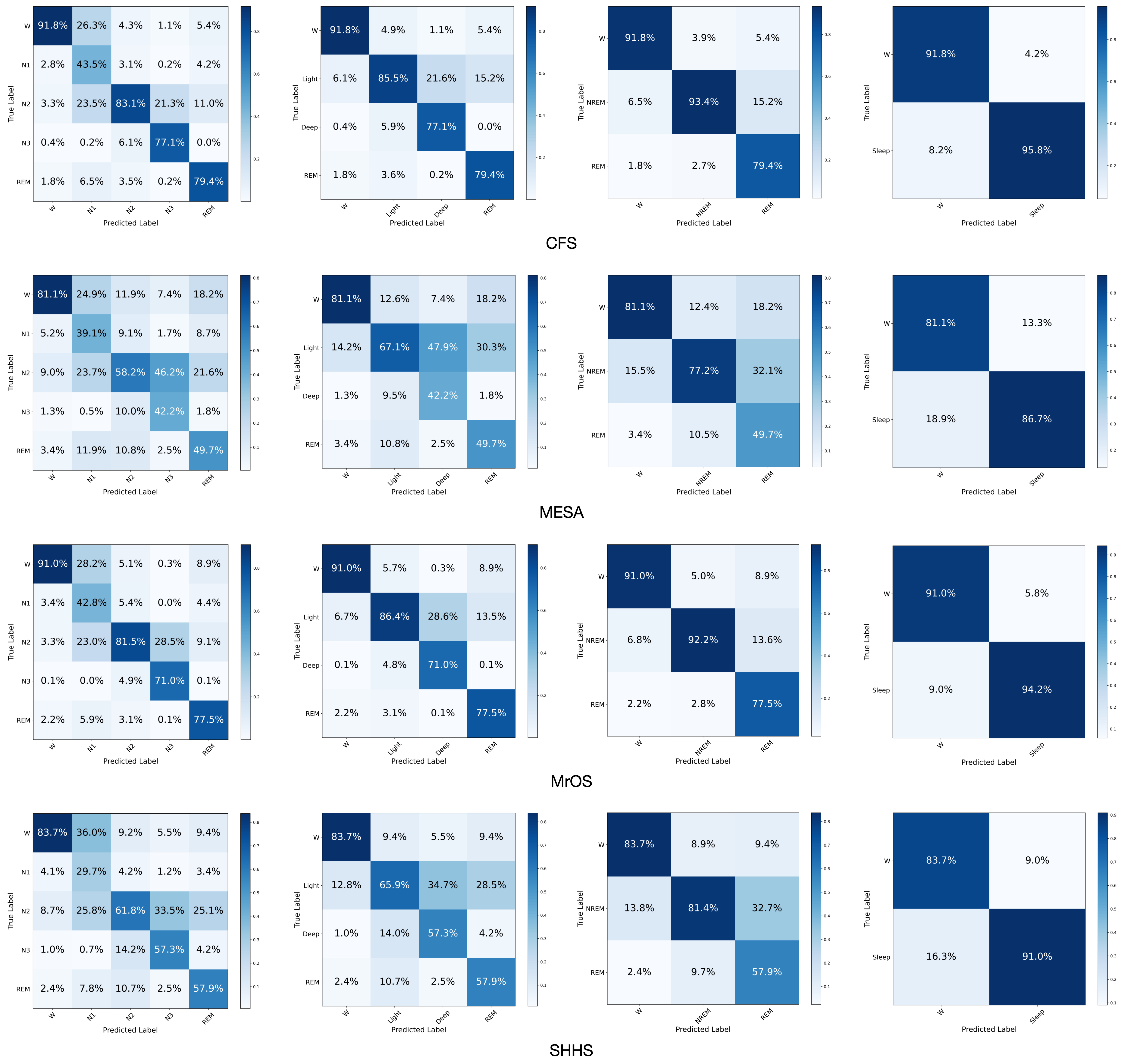}
    \captionof{figure}{Confusion matrices for cohort-specific sleep staging models.}
    \label{fig:appendixA_confusion}
\end{minipage}

\vspace{1em}

\noindent\begin{minipage}{\linewidth}
    \centering
    \includegraphics[width=0.9\textwidth]{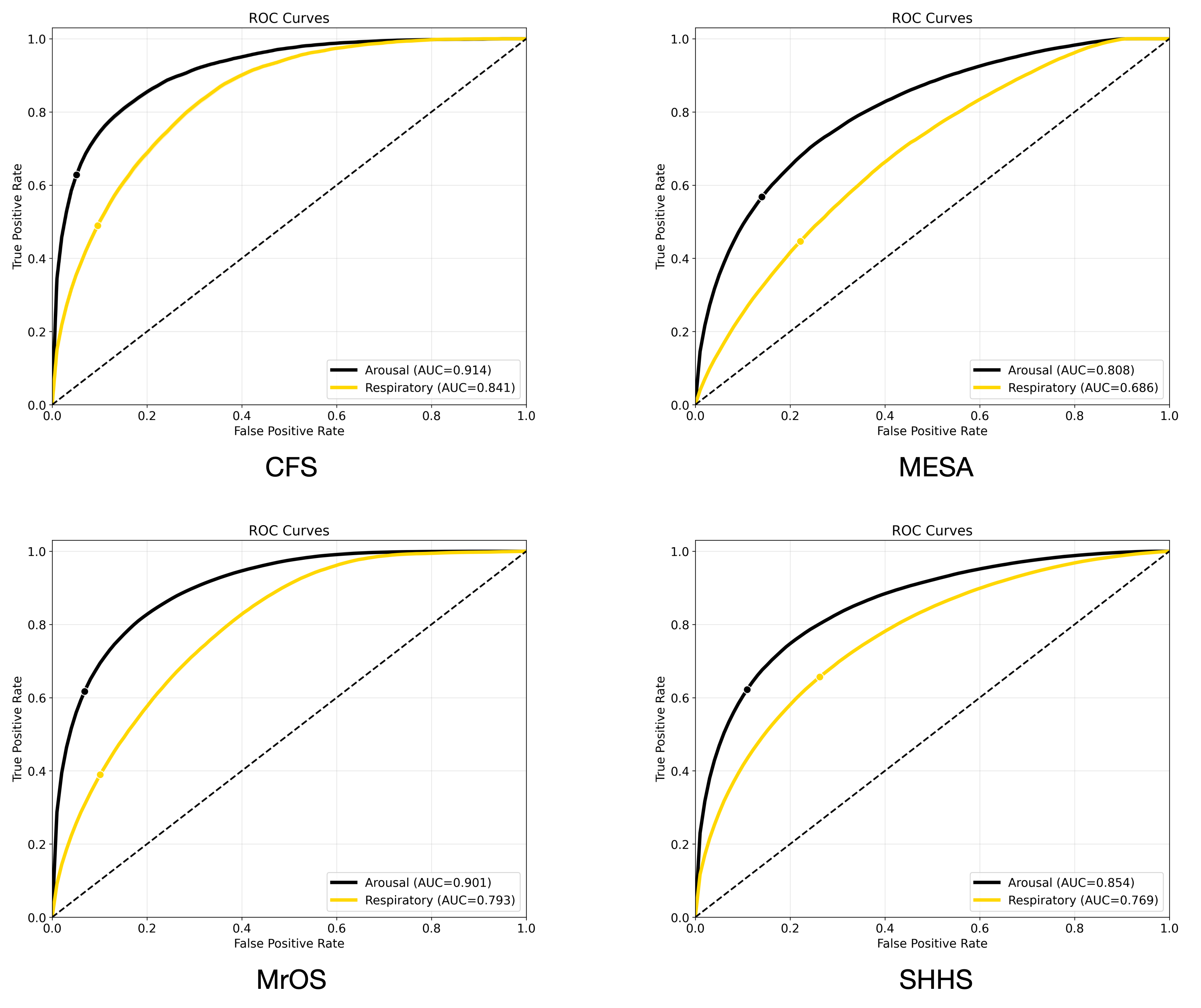}
    \captionof{figure}{ROC curves for cohort-specific sleep disturbance detection models.}
    \label{fig:appendixA_roc}
\end{minipage}


\newpage
\section{Ablation Study}
\label{app:ablation}
{
\FloatBarrier

\noindent
To quantify the contribution of key design choices under a controlled computational budget, all ablation experiments were conducted on the CFS cohort, while keeping the training and evaluation protocol consistent with the main experiments. We investigate (i) feature fusion, (ii) model composition within the two-stage framework, and (iii) the multi-frame window length used by the Transformer-based temporal refinement module.

\subsection{Feature fusion}
\label{app:ablation:fusion}
\FloatBarrier

\begin{table}[H]
  \caption{Ablation on feature fusion for sleep staging on CFS. We use the two-stage model, and the Transformer refinement stage adopts multi-frame concatenation with a fixed window size of 15. The comparison evaluates time-domain input only versus feature-level fusion of time-domain representations with frequency/time--frequency descriptors. Feature fusion yields higher \(\kappa\) and improves overall metrics.}
  \label{tab:ablation_fusion_staging}
  \centering
  \small
  \setlength{\tabcolsep}{6pt}
  \begin{adjustbox}{max width=\linewidth}
    \begin{tabular}{lcccc}
      \toprule
      \textbf{Feature Fusion} & \textbf{Accuracy} $\uparrow$ & \textbf{Weighted F1} $\uparrow$ & $\boldsymbol{\kappa}$ $\uparrow$ & \textbf{Macro AUC} $\uparrow$ \\
      \midrule
      With Fusion    & 0.842 & 0.835 & 0.776 & 0.959 \\
      Without Fusion & 0.823 & 0.819 & 0.748 & 0.949 \\
      \bottomrule
    \end{tabular}
  \end{adjustbox}
\end{table}

\FloatBarrier

\begin{table}[H]
  \caption{Ablation on feature fusion for multi-task sleep disturbance detection on CFS (arousal and respiratory events). The two-stage model is used, and the Transformer refinement stage adopts a fixed multi-frame window size of 15. Feature fusion improves AUC for both tasks and shows a more pronounced benefit for sleep staging (Table~\ref{tab:ablation_fusion_staging}).}
  \label{tab:ablation_fusion_events}
  \centering
  \scriptsize
  \setlength{\tabcolsep}{6pt}
  \begin{adjustbox}{max width=\linewidth}
    \begin{tabular}{llcccccc}
      \toprule
      \textbf{Feature Fusion} & \textbf{Task} &
      \textbf{Accuracy} $\uparrow$ & \textbf{Precision} $\uparrow$ & \textbf{Recall} $\uparrow$ & \textbf{Specificity} $\uparrow$ & \textbf{F1-Score} $\uparrow$ & \textbf{AUC} $\uparrow$ \\
      \midrule
      \multirow{2}{*}{With Fusion}
      & Arousal     & 0.911 & 0.618 & 0.628 & 0.949 & 0.623 & 0.914 \\
      & Respiratory & 0.868 & 0.325 & 0.490 & 0.904 & 0.391 & 0.841 \\
      \midrule
      \multirow{2}{*}{Without Fusion}
      & Arousal     & 0.900 & 0.569 & 0.629 & 0.936 & 0.597 & 0.902 \\
      & Respiratory & 0.847 & 0.365 & 0.543 & 0.884 & 0.437 & 0.835 \\
      \bottomrule
    \end{tabular}
  \end{adjustbox}
\end{table}

\FloatBarrier

\subsection{Model comparison}
\label{app:ablation:model}
\FloatBarrier

\begin{table}[H]
  \caption{Model comparison for sleep staging on CFS within the proposed two-stage framework. All settings use feature fusion. We compare (i) a stage-1 CNN-only model (Net1D), (ii) a Transformer-only model without the CNN feature extractor, and (iii) the full two-stage model (Net1D + Transformer). For this ablation, the Transformer input window size is fixed to 1 (no multi-frame concatenation).}
  \label{tab:ablation_model_staging}
  \centering
  \small
  \setlength{\tabcolsep}{6pt}
  \begin{adjustbox}{max width=\linewidth}
    \begin{tabular}{lcccc}
      \toprule
      \textbf{Model} & \textbf{Accuracy} $\uparrow$ & \textbf{Weighted F1} $\uparrow$ & $\boldsymbol{\kappa}$ $\uparrow$ & \textbf{Macro AUC} $\uparrow$ \\
      \midrule
      Net1D                 & 0.735 & 0.728 & 0.630 & 0.912 \\
      Transformer               & 0.756 & 0.743 & 0.651 & 0.912 \\
      Net1D + Transformer   & 0.791 & 0.780 & 0.703 & 0.924 \\
      \bottomrule
    \end{tabular}
  \end{adjustbox}
\end{table}

\FloatBarrier

\begin{table}[H]
  \caption{Model comparison for multi-task sleep event detection on CFS (arousal and respiratory events). All settings use feature fusion. We compare the stage-1 CNN-only model, a Transformer-only model, and the full two-stage model. For this ablation, the Transformer input window size is fixed to 1 (no multi-frame concatenation).}
  \label{tab:ablation_model_events}
  \centering
  \scriptsize
  \setlength{\tabcolsep}{3.5pt}
  \begin{adjustbox}{max width=\linewidth}
    \begin{tabular}{llcccccc}
      \toprule
      \textbf{Model} & \textbf{Task} &
      \textbf{Accuracy} $\uparrow$ & \textbf{Precision} $\uparrow$ & \textbf{Recall} $\uparrow$ & \textbf{Specificity} $\uparrow$ & \textbf{F1-Score} $\uparrow$ & \textbf{AUC} $\uparrow$ \\
      \midrule
      \multirow{2}{*}{CNN Net1D}
      & Arousal     & 0.884 & 0.510 & 0.563 & 0.927 & 0.535 & 0.868 \\
      & Respiratory & 0.741 & 0.192 & 0.532 & 0.763 & 0.283 & 0.734 \\
      \midrule
      \multirow{2}{*}{Transformer}
      & Arousal     & 0.705 & 0.197 & 0.487 & 0.734 & 0.281 & 0.662 \\
      & Respiratory & 0.530 & 0.133 & 0.711 & 0.511 & 0.225 & 0.647 \\
      \midrule
      \multirow{2}{*}{CNN Net1D + Transformer}
      & Arousal     & 0.887 & 0.514 & 0.602 & 0.924 & 0.555 & 0.878 \\
      & Respiratory & 0.816 & 0.302 & 0.522 & 0.852 & 0.383 & 0.793 \\
      \bottomrule
    \end{tabular}
  \end{adjustbox}
\end{table}

\FloatBarrier

\subsection{Multi-frame window size}
\label{app:ablation:window}
\FloatBarrier

\begin{table}[H]
  \caption{Ablation on the multi-frame window length used by the Transformer-based refinement module for sleep staging on CFS. The two-stage model with feature fusion is used, and only the window size is varied from 1 to 21. Performance generally increases as the window length grows, and a window size of 15 provides a strong accuracy--efficiency trade-off without introducing substantial additional computation.}
  \label{tab:ablation_window_staging}
  \centering
  \small
  \setlength{\tabcolsep}{5.5pt}
  \begin{adjustbox}{max width=\linewidth}
    \begin{tabular}{ccccc}
      \toprule
      \textbf{Window size} & \textbf{Accuracy} $\uparrow$ & \textbf{Weighted F1} $\uparrow$ & $\boldsymbol{\kappa}$ $\uparrow$ & \textbf{Macro AUC} $\uparrow$ \\
      \midrule
      1  & 0.791 & 0.780 & 0.703 & 0.924 \\
      3  & 0.824 & 0.814 & 0.750 & 0.945 \\
      5  & 0.833 & 0.824 & 0.763 & 0.951 \\
      7  & 0.838 & 0.830 & 0.771 & 0.955 \\
      9  & 0.841 & 0.834 & 0.775 & 0.958 \\
      11 & 0.836 & 0.830 & 0.768 & 0.956 \\
      13 & 0.839 & 0.831 & 0.772 & 0.956 \\
      15 & 0.842 & 0.835 & 0.776 & 0.959 \\
      17 & 0.842 & 0.836 & 0.775 & 0.958 \\
      19 & 0.842 & 0.837 & 0.776 & 0.959 \\
      21 & 0.838 & 0.833 & 0.771 & 0.957 \\
      \bottomrule
    \end{tabular}
  \end{adjustbox}
\end{table}

\FloatBarrier

\begin{table}[H]
  \caption{Ablation on the multi-frame window length for multi-task sleep disturbance detection on CFS. The two-stage model with feature fusion is used, and only the Transformer window size is varied from 1 to 21. Overall performance improves as the window length increases, and a window size of 15 provides strong discrimination with limited additional computational overhead.}
  \label{tab:ablation_window_events}
  \centering
  \scriptsize
  \setlength{\tabcolsep}{4pt}
  \begin{adjustbox}{max width=\linewidth}
    \begin{tabular}{clcccccc}
      \toprule
      \textbf{Window size} & \textbf{Task} &
      \textbf{Accuracy} $\uparrow$ & \textbf{Precision} $\uparrow$ & \textbf{Recall} $\uparrow$ & \textbf{Specificity} $\uparrow$ & \textbf{F1-Score} $\uparrow$ & \textbf{AUC} $\uparrow$ \\
      \midrule
      \multirow{2}{*}{1}
      & Arousal     & 0.887 & 0.514 & 0.602 & 0.924 & 0.555 & 0.878 \\
      & Respiratory & 0.816 & 0.302 & 0.522 & 0.852 & 0.383 & 0.793 \\
      \midrule
      \multirow{2}{*}{3}
      & Arousal     & 0.897 & 0.556 & 0.625 & 0.934 & 0.588 & 0.899 \\
      & Respiratory & 0.838 & 0.344 & 0.534 & 0.875 & 0.418 & 0.824 \\
      \midrule
      \multirow{2}{*}{5}
      & Arousal     & 0.903 & 0.583 & 0.595 & 0.944 & 0.589 & 0.898 \\
      & Respiratory & 0.845 & 0.359 & 0.529 & 0.884 & 0.428 & 0.827 \\
      \midrule
      \multirow{2}{*}{7}
      & Arousal     & 0.902 & 0.578 & 0.602 & 0.942 & 0.590 & 0.898 \\
      & Respiratory & 0.849 & 0.366 & 0.525 & 0.889 & 0.432 & 0.830 \\
      \midrule
      \multirow{2}{*}{9}
      & Arousal     & 0.898 & 0.563 & 0.606 & 0.937 & 0.584 & 0.895 \\
      & Respiratory & 0.843 & 0.357 & 0.544 & 0.880 & 0.431 & 0.828 \\
      \midrule
      \multirow{2}{*}{11}
      & Arousal     & 0.898 & 0.560 & 0.614 & 0.936 & 0.586 & 0.896 \\
      & Respiratory & 0.838 & 0.349 & 0.563 & 0.871 & 0.431 & 0.830 \\
      \midrule
      \multirow{2}{*}{13}
      & Arousal     & 0.905 & 0.592 & 0.622 & 0.943 & 0.607 & 0.906 \\
      & Respiratory & 0.850 & 0.373 & 0.543 & 0.888 & 0.442 & 0.840 \\
      \midrule
      \multirow{2}{*}{15}
      & Arousal     & 0.911 & 0.618 & 0.628 & 0.949 & 0.623 & 0.914 \\
      & Respiratory & 0.868 & 0.325 & 0.490 & 0.904 & 0.391 & 0.841 \\
      \midrule
      \multirow{2}{*}{17}
      & Arousal     & 0.902 & 0.578 & 0.628 & 0.939 & 0.602 & 0.904 \\
      & Respiratory & 0.854 & 0.379 & 0.517 & 0.896 & 0.438 & 0.837 \\
      \midrule
      \multirow{2}{*}{19}
      & Arousal     & 0.906 & 0.597 & 0.613 & 0.945 & 0.605 & 0.904 \\
      & Respiratory & 0.857 & 0.387 & 0.526 & 0.897 & 0.446 & 0.840 \\
      \midrule
      \multirow{2}{*}{21}
      & Arousal     & 0.900 & 0.568 & 0.618 & 0.937 & 0.592 & 0.898 \\
      & Respiratory & 0.852 & 0.374 & 0.520 & 0.893 & 0.435 & 0.834 \\
      \bottomrule
    \end{tabular}
  \end{adjustbox}
\end{table}

\FloatBarrier
}

\newpage
\section{Comparison of Existing Studies}
\label{app:comparison}

\begin{table}[h!]
\centering
\setlength{\tabcolsep}{4pt}
\renewcommand{\arraystretch}{1.15}
\caption{\textbf{Representative single-lead ECG-based sleep-staging studies and their self-reported metrics.}
Reported values are taken from the original publications and correspond to each study's internal test-set evaluation under the label set indicated (3-/4-/5-class). $N$ denotes the number of PSG recordings unless otherwise specified. Because cohorts, label definitions, and evaluation protocols differ across studies, this table is provided for qualitative context rather than head-to-head benchmarking. Missing entries indicate that the corresponding metric was not reported in the original publication.}
\label{tab:app_ecg_lit}

\begin{tabularx}{\textwidth}{
>{\raggedright\arraybackslash}p{3.3cm}
c
>{\raggedright\arraybackslash}X
r
c
c
}
\toprule
\textbf{Study} & \textbf{Classes} & \textbf{Dataset(s)} & \textbf{$N$} & \textbf{$\kappa$} & \textbf{Acc} \\
\midrule

Wei \textit{et al.} (2018)~\cite{wei2018research} & 3 & MIT--BIH Polysomnographic Database~\cite{goldberger2000components} & 18 & 0.560 & 0.770 \\
Urtnasan \textit{et al.} (2022)~\cite{urtnasan2022deep} & 3 & Private dataset & 112 & -- & 0.860 \\
\addlinespace

Li \textit{et al.} (2018)~\cite{li2018deep} & 4 & MIT--BIH Polysomnographic Database~\cite{goldberger2000components} & 18 & 0.540 & 0.756 \\
Wei \textit{et al.} (2019)~\cite{wei2019multi} & 4 & Xijing Hospital dataset (FMMU) & 238 & 0.550 & 0.778 \\
Radha \textit{et al.} (2019)~\cite{radha2019sleep} & 4 & Siesta database~\cite{klosh2001siesta} & 588 & 0.610 & 0.770 \\
\addlinespace

Sun \textit{et al.} (2020)~\cite{sun2020sleep} & 5 & MGH Cohort & 8682 & 0.490 & -- \\
\addlinespace

\textbf{Holter-to-Sleep (this study)} & 5 & MESA~\cite{chen2015racial} + MrOS~\cite{blackwell2011associations} + SHHS~\cite{quan1997sleep} & 9730 & 0.670 & 0.773 \\
\bottomrule
\end{tabularx}
\end{table}


\newpage
\section{Supplementary Methods}
\label{app:ablation}

\subsection{Input representation and embedding dimensionality}
Each 30-s epoch was represented by (i) a time-domain single-lead ECG segment (stored as a 1D vector of length 3000 with an explicit channel dimension) and (ii) an epoch-aligned frequency-domain feature vector (39 dimensions). The epoch-level feature extractor was a Net1D-style architecture with attention-based aggregation for the time-domain backbone output, and a lightweight feed-forward mapping with attention weighting for the frequency-domain features. The final epoch embedding was obtained by concatenating a 1024-dimensional time-domain representation and a 128-dimensional frequency-domain representation, resulting in a 1152-dimensional embedding per epoch.

\subsection{Transformer architectures and context-window definition}
\paragraph{Sleep-staging Transformer (sequence-level).}
The sleep-staging Transformer used a batch-first Transformer Encoder with learnable positional embeddings. For each epoch, we formed a local context window of length $W=15$ (7 preceding epochs, the current epoch, and 7 following epochs). The 15 epoch embeddings (each 1152-dimensional) within the window were concatenated to form the input vector of a single token, yielding a token input dimension of $15\times1152$. The model projected tokens to hidden size $d=512$, added positional embeddings, and applied a $L=3$-layer Transformer Encoder ($h=8$ heads, feed-forward dimension $4d$, dropout 0.1). The classifier produced 5-class sleep-stage logits for each token in the sequence. Because the number of epochs varies across recordings, sequences were padded within a batch; loss computation ignored padded positions via an \texttt{ignore\_index} mask.

\paragraph{Sleep-event Transformer (window-level).}
For event detection, each local window ($W=15$) was treated as an individual sample with input shape $(B,W,1152)$. After linear projection to $d=512$, a $L=4$-layer Transformer Encoder ($h=8$, dropout 0.1) generated contextualized hidden states. The center-token representation was used as a shared embedding for two parallel binary classifiers, producing arousal and respiratory logits, respectively.

\subsection{Training configuration and hyperparameters}
\paragraph*{Sleep-staging CNN (epoch-level).}
We used Adam with learning rate $1\times10^{-3}$ and weight decay $1\times10^{-5}$, trained for 50 epochs with batch size 256. A ReduceLROnPlateau scheduler was applied (patience=10, factor=0.3), and early stopping was performed based on validation loss (patience=5). Training supported distributed data parallelism (DDP) and automatic mixed precision (AMP).

\paragraph{Sleep-staging Transformer.}
We used Adam with learning rate $1\times10^{-4}$, batch size 64, and up to 100 epochs, with early stopping based on validation loss (patience=5). The context window length was $W=15$ with stride 1. Cross-entropy loss was computed with padded labels ignored by \texttt{ignore\_index}.

\paragraph{Multitask sleep-event CNN.}
The multitask CNN shared a Net1D encoder and used two binary classification heads. We used AdamW with learning rate $1\times10^{-3}$ and weight decay $1\times10^{-5}$, trained for 50 epochs with batch size 128. ReduceLROnPlateau was applied (patience=5, factor=0.3) with early stopping (patience=3). During training, small additive Gaussian noise (standard deviation 0.01) was used for augmentation. Training supported DDP and AMP.

\paragraph{Sleep-event Transformer.}
We used Adam with learning rate $3\times10^{-4}$, batch size 16, and up to 200 epochs, with early stopping based on validation loss (patience=6). The context window length was $W=15$ with stride 1, and logits were computed from the center-token representation using two parallel binary heads.

\subsection{Split unit and generation of file lists}
Train/validation/test splits were performed at the recording-file level (one \texttt{.npz} file per overnight recording). All \texttt{.npz} paths were collected recursively from the specified directories and randomly split with a fixed seed. We first assigned 60\% of files to training; the remaining 40\% was equally divided into validation and test sets (20\%/20\%). The resulting file lists were written to \texttt{train\_paths.txt}, \texttt{val\_paths.txt}, and \texttt{test\_paths.txt}. All training and evaluation scripts used these lists as the sole data input, ensuring reproducible splits and results.

\subsection{Threshold selection for binary tasks}
For binary event detection, the model outputs were converted to positive-class probabilities $p$ via softmax. The decision threshold $t$ was determined on the validation set by grid search: we evaluated 100 equally spaced candidates in $t\in[0.01,0.99]$ and selected the threshold that maximized the F1-score. The selected threshold was then fixed for test evaluation, reporting Accuracy, Precision, Recall, Specificity, F1-score, and AUC.

\subsection{Bootstrap unit and confidence intervals}
We used nonparametric bootstrap to quantify metric uncertainty. The resampling unit was the epoch (30-s segment): the evaluation set was flattened into an epoch-level sample list, which was resampled with replacement for 1000 replicates. For each replicate, the target metric was computed, and the 2.5th and 97.5th percentiles of the bootstrap distribution were used to construct a 95\% confidence interval; the bootstrap mean was reported as the point estimate.

\clearpage
\section{List of sleep and Holter-grade cardiac metrics}
\label{app:metric_list}

\noindent
For reproducibility and reference, this appendix summarizes the sleep-related metrics (16 items) and
Holter-grade cardiac metrics (32 items) used in this study.

\begin{table}[H]
  \centering
  \small
  \setlength{\tabcolsep}{10pt}
  \renewcommand{\arraystretch}{1.25}
  \caption{Summary of sleep-related metrics and Holter-grade cardiac metrics used in this study.}
  \label{tab:metric_list}
  \begin{tabularx}{\linewidth}{>{\raggedright\arraybackslash}p{0.44\linewidth} >{\raggedright\arraybackslash}p{0.52\linewidth}}
    \toprule
    \textbf{Sleep metrics (16)} & \textbf{Holter-grade cardiac metrics (32)} \\
    \midrule
    AHI (events\,h$^{-1}$) \newline
    ArI (events\,h$^{-1}$) \newline
    N1 / TST (\%) \newline
    N2 / TST (\%) \newline
    N3 / TST (\%) \newline
    NREM / TST (\%) \newline
    REM / TST (\%) \newline
    SE (\%) \newline
    SL (min) \newline
    TIB (min) \newline
    TST (min) \newline
    TTSP (min) \newline
    WASO (min) \newline
    Wake / TIB (\%) \newline
    ODI 3\% (events\,h$^{-1}$) \newline
    ODI 4\% (events\,h$^{-1}$)
    &
    Total\_valid\_beats \newline
    Avg\_HR \newline
    Min\_HR \newline
    Max\_HR \newline
    Snt\_max\_beat\_count \newline
    Snt\_duration \newline
    Snb\_max\_beat\_count \newline
    Snb\_duration \newline
    Total\_PAC \newline
    Single\_PAC \newline
    Paired\_PAC \newline
    Bigeminy\_PAC \newline
    Triad\_PAC \newline
    Total\_PVC \newline
    Single\_PVC \newline
    Paired\_PVC \newline
    Bigeminy\_PVC \newline
    Triad\_PVC \newline
    Total\_VT \newline
    Total\_SVT \newline
    SDNN \newline
    SDANN \newline
    SDANNIndex \newline
    RMSSD \newline
    pNN50 \newline
    HRV\_tri \newline
    LF \newline
    HF \newline
    LFNU \newline
    HFNU \newline
    LF/HF \newline
    AF\_duration \\
    \bottomrule
  \end{tabularx}
\end{table}

\clearpage
\section{Handcrafted frequency and time--frequency features}
\label{app:handcrafted_features}

\begin{table}[H]
    \centering
    \caption{Handcrafted spectral and time--frequency descriptors (39 features) used for feature-level fusion.}
    \label{tab:frequency_features}
    \renewcommand{\arraystretch}{1.5} 
    \setlength{\tabcolsep}{10pt} 

    \begin{tabular}{l p{10cm}}
        \toprule
        \textbf{Frequency Domain Features} & \textbf{Time-Frequency Features} \\
        \midrule
        1. Delta Power & \textbf{Energy Features}: Energy calculated for each wavelet decomposition level (5 levels, first-level coefficients for each level) \\
        2. Theta Power & \textbf{Statistical Features}: Statistical features computed for each wavelet decomposition level (5 $\times$ 5 matrix, 5 features per decomposition level), including: \\
        3. Alpha Power & 1. Mean \\
        4. Total Power & 2. Standard Deviation \\
        5. Peak Frequency & 3. Variance \\
        6. Spectral Centroid & 4. Maximum Value \\
        7. Spectral Bandwidth & 5. Minimum Value \\
        8. Spectral Flatness &  \\
        9. Spectral Entropy &  \\
        \bottomrule
    \end{tabular}
\end{table}

\clearpage
\section{Detailed tables for joint sleep--Holter stratified analyses}
\label{app:joint_tables}

\noindent
This appendix reports the complete metric tables used in the joint sleep--Holter stratified analyses.
Values are presented as mean $\pm$ standard deviation.
P-values are from two-sided statistical tests; extremely small P-values are reported as ``<0.001'',
otherwise rounded to three decimals.

\subsection{Frequent PVC vs non-frequent PVC}
\label{app:joint_tables_pvc}

\begin{table}[H]
  \centering
  \small
  \setlength{\tabcolsep}{6pt}
  \renewcommand{\arraystretch}{1.15}
  \caption{Full metric comparison between frequent PVC and non-frequent PVC strata. Values are mean $\pm$ SD.}
  \label{tab:joint_pvc_full}
  \begin{adjustbox}{max width=\linewidth}
    \begin{tabular}{lccc}
      \toprule
      \textbf{Metric} &
      \textbf{Frequent PVC (n=1,248)} &
      \textbf{Non-frequent PVC (n=9,191)} &
      \textbf{P-value} \\
      \midrule
      Age (years)                 & 69.83 $\pm$ 14.31 & 66.29 $\pm$ 12.84 & <0.001 \\
      BMI (kg\,m$^{-2}$)                 & 28.51 $\pm$ 5.66  & 28.25 $\pm$ 5.29  & 0.250 \\
      AHI (events\,h$^{-1}$)    & 21.48 $\pm$ 18.03 & 19.11 $\pm$ 16.86 & <0.001 \\
      ArI (events\,h$^{-1}$)                       & 31.36 $\pm$ 18.09 & 29.13 $\pm$ 15.38 & <0.001 \\
      N1 \% of TST              & 7.87 $\pm$ 7.10   & 7.19 $\pm$ 6.23   & <0.001 \\
      N2 \% of TST              & 59.56 $\pm$ 11.85 & 58.01 $\pm$ 11.33 & <0.001 \\
      N3 \% of TST              & 13.93 $\pm$ 11.47 & 15.22 $\pm$ 11.44 & <0.001 \\
      NREM \% of TST            & 81.36 $\pm$ 7.17  & 80.42 $\pm$ 6.57  & <0.001 \\
      REM \% of TST             & 18.64 $\pm$ 7.17  & 19.58 $\pm$ 6.57  & <0.001 \\
      SE (\%)                        & 62.85 $\pm$ 15.07 & 68.70 $\pm$ 14.25 & <0.001 \\
      SL (min)                  & 68.59 $\pm$ 66.18 & 50.96 $\pm$ 55.26 & <0.001 \\
      TIB (min)                 & 553.12 $\pm$ 81.75& 529.91 $\pm$ 78.58& <0.001 \\
      TST (min)                 & 342.80 $\pm$ 77.32& 358.47 $\pm$ 66.20& <0.001 \\
      TTSP (min)                & 443.97 $\pm$ 82.54& 438.50 $\pm$ 69.40& 0.002 \\
      WASO (min)                & 101.18 $\pm$ 71.44& 80.02 $\pm$ 57.92 & <0.001 \\
      Wake \% of TIB            & 37.15 $\pm$ 15.07 & 31.30 $\pm$ 14.25 & <0.001 \\
      ODI 3\% (events\,h$^{-1}$)                   & 7.58 $\pm$ 10.75  & 6.20 $\pm$ 9.59   & <0.001 \\
      ODI 4\% (events\,h$^{-1}$)                   & 4.38 $\pm$ 7.62   & 3.61 $\pm$ 7.15   & <0.001 \\
      \bottomrule
    \end{tabular}
  \end{adjustbox}
\end{table}

\subsection{Frequent PAC vs non-frequent PAC}
\label{app:joint_tables_pac}

\begin{table}[H]
  \centering
  \small
  \setlength{\tabcolsep}{6pt}
  \renewcommand{\arraystretch}{1.15}
  \caption{Full metric comparison between frequent PAC and non-frequent PAC strata. Values are mean $\pm$ SD.}
  \label{tab:joint_pac_full}
  \begin{adjustbox}{max width=\linewidth}
    \begin{tabular}{lccc}
      \toprule
      \textbf{Metric} &
      \textbf{Frequent PAC (n=2,022)} &
      \textbf{Non-frequent PAC (n=8,417)} &
      \textbf{P-value} \\
      \midrule
      Age (years)                 & 68.96 $\pm$ 16.03 & 66.17 $\pm$ 12.20 & <0.001 \\
      BMI (kg\,m$^{-2}$)                & 28.08 $\pm$ 5.26  & 28.33 $\pm$ 5.35  & 0.074 \\
      AHI (events\,h$^{-1}$)    & 22.09 $\pm$ 19.38 & 18.74 $\pm$ 16.34 & <0.001 \\
      ArI (events\,h$^{-1}$)                       & 32.35 $\pm$ 18.85 & 28.69 $\pm$ 14.81 & <0.001 \\
      N1 \% of TST              & 8.00 $\pm$ 6.89   & 7.10 $\pm$ 6.19   & <0.001 \\
      N2 \% of TST              & 59.25 $\pm$ 11.68 & 57.94 $\pm$ 11.32 & <0.001 \\
      N3 \% of TST              & 14.14 $\pm$ 11.75 & 15.29 $\pm$ 11.37 & <0.001 \\
      NREM \% of TST            & 81.39 $\pm$ 6.95  & 80.33 $\pm$ 6.56  & <0.001 \\
      REM \% of TST             & 18.61 $\pm$ 6.95  & 19.67 $\pm$ 6.56  & <0.001 \\
      SE (\%)                        & 63.67 $\pm$ 14.86 & 69.04 $\pm$ 14.18 & <0.001 \\
      SL (min)                  & 64.31 $\pm$ 63.75 & 50.37 $\pm$ 54.86 & <0.001 \\
      TIB (min)                 & 548.36 $\pm$ 81.56& 528.92 $\pm$ 78.31& <0.001 \\
      TST (min)                 & 344.11 $\pm$ 74.81& 359.60 $\pm$ 65.67& <0.001 \\
      TTSP (min)                & 442.30 $\pm$ 79.98& 438.39 $\pm$ 68.81& 0.006 \\
      WASO (min)                & 98.19 $\pm$ 69.37 & 78.79 $\pm$ 57.00 & <0.001 \\
      Wake \% of TIB            & 36.33 $\pm$ 14.86 & 30.96 $\pm$ 14.18 & <0.001 \\
      ODI 3\% (events\,h$^{-1}$)                   & 7.86 $\pm$ 11.61  & 6.00 $\pm$ 9.21   & <0.001 \\
      ODI 4\% (events\,h$^{-1}$)                  & 4.61 $\pm$ 8.46   & 3.49 $\pm$ 6.86   & <0.001 \\
      \bottomrule
    \end{tabular}
  \end{adjustbox}
\end{table}

\subsection{Atrial fibrillation during sleep: AF present vs absent}
\label{app:joint_tables_af}

\begin{table}[H]
  \centering
  \small
  \setlength{\tabcolsep}{6pt}
  \renewcommand{\arraystretch}{1.15}
  \caption{Full metric comparison between AF present and AF absent strata during sleep. Values are mean $\pm$ SD.}
  \label{tab:joint_af_full}
  \begin{adjustbox}{max width=\linewidth}
    \begin{tabular}{lccc}
      \toprule
      \textbf{Metric} &
      \textbf{AF present (n=1,162)} &
      \textbf{AF absent (n=9,277)} &
      \textbf{P-value} \\
      \midrule
      Age (years)                 & 65.62 $\pm$ 19.41 & 66.85 $\pm$ 12.03 & <0.001 \\
      BMI (kg\,m$^{-2}$)                 & 28.13 $\pm$ 5.34  & 28.30 $\pm$ 5.33  & 0.535 \\
      AHI (events\,h$^{-1}$)    & 20.87 $\pm$ 18.79 & 19.21 $\pm$ 16.77 & 0.104 \\
      ArI (events\,h$^{-1}$)                       & 30.69 $\pm$ 18.20 & 29.24 $\pm$ 15.40 & 0.204 \\
      N1 \% of TST              & 7.22 $\pm$ 6.22   & 7.28 $\pm$ 6.36   & 0.572 \\
      N2 \% of TST              & 59.11 $\pm$ 11.90 & 58.08 $\pm$ 11.34 & <0.001 \\
      N3 \% of TST              & 14.77 $\pm$ 12.40 & 15.10 $\pm$ 11.33 & 0.019 \\
      NREM \% of TST            & 81.10 $\pm$ 6.85  & 80.46 $\pm$ 6.62  & <0.001 \\
      REM \% of TST             & 18.90 $\pm$ 6.85  & 19.54 $\pm$ 6.62  & <0.001 \\
      SE (\%)                        & 65.73 $\pm$ 13.87 & 68.28 $\pm$ 14.52 & <0.001 \\
      SL (min)                  & 63.98 $\pm$ 63.25 & 51.69 $\pm$ 55.96 & <0.001 \\
      TIB (min)                 & 548.11 $\pm$ 79.79& 530.74 $\pm$ 79.05& <0.001 \\
      TST (min)                 & 355.18 $\pm$ 68.35& 356.78 $\pm$ 67.74& 0.162 \\
      TTSP (min)                & 448.31 $\pm$ 74.01& 437.99 $\pm$ 70.67& <0.001 \\
      WASO (min)                & 93.12 $\pm$ 68.13 & 81.22 $\pm$ 58.86 & <0.001 \\
      Wake \% of TIB            & 34.27 $\pm$ 13.87 & 31.72 $\pm$ 14.52 & <0.001 \\
      ODI 3\% (events\,h$^{-1}$)                   & 7.56 $\pm$ 11.51  & 6.21 $\pm$ 9.49   & <0.001 \\
      ODI 4\% (events\,h$^{-1}$)                   & 4.45 $\pm$ 8.42   & 3.61 $\pm$ 7.03   & <0.001 \\
      \bottomrule
    \end{tabular}
  \end{adjustbox}
\end{table}

\subsection{Ventricular tachycardia during sleep: VT present vs absent}
\label{app:joint_tables_vt}

\begin{table}[H]
  \centering
  \small
  \setlength{\tabcolsep}{6pt}
  \renewcommand{\arraystretch}{1.15}
  \caption{Full metric comparison between VT present and VT absent strata during sleep. Values are mean $\pm$ SD.}
  \label{tab:joint_vt_full}
  \begin{adjustbox}{max width=\linewidth}
    \begin{tabular}{lccc}
      \toprule
      \textbf{Metric} &
      \textbf{VT present (n=489)} &
      \textbf{VT absent (n=9,950)} &
      \textbf{P-value} \\
      \midrule
      Age (years)                 & 69.93 $\pm$ 15.16 & 66.56 $\pm$ 12.94 & <0.001 \\
      BMI (kg\,m$^{-2}$)                 & 28.66 $\pm$ 5.91  & 28.26 $\pm$ 5.30  & 0.468 \\
      AHI (events\,h$^{-1}$)    & 20.99 $\pm$ 16.74 & 19.32 $\pm$ 17.03 & 0.001 \\
      ArI (events\,h$^{-1}$)                       & 30.00 $\pm$ 16.49 & 29.37 $\pm$ 15.71 & 0.388 \\
      N1 \% of TST              & 7.18 $\pm$ 5.73   & 7.28 $\pm$ 6.37   & 0.617 \\
      N2 \% of TST              & 60.24 $\pm$ 11.70 & 58.10 $\pm$ 11.38 & <0.001 \\
      N3 \% of TST              & 13.57 $\pm$ 11.40 & 15.14 $\pm$ 11.45 & <0.001 \\
      NREM \% of TST            & 80.99 $\pm$ 6.88  & 80.51 $\pm$ 6.64  & 0.091 \\
      REM \% of TST             & 19.01 $\pm$ 6.88  & 19.49 $\pm$ 6.64  & 0.091 \\
      SE (\%)                        & 63.26 $\pm$ 14.32 & 68.23 $\pm$ 14.44 & <0.001 \\
      SL (min)                  & 74.05 $\pm$ 66.62 & 52.04 $\pm$ 56.24 & <0.001 \\
      TIB (min)                 & 564.65 $\pm$ 81.98& 531.11 $\pm$ 78.86& <0.001 \\
      TST (min)                 & 352.44 $\pm$ 75.06& 356.80 $\pm$ 67.43& 0.217 \\
      TTSP (min)                & 454.35 $\pm$ 78.17& 438.40 $\pm$ 70.68& <0.001 \\
      WASO (min)                & 101.91 $\pm$ 69.93& 81.60 $\pm$ 59.40 & <0.001 \\
      Wake \% of TIB            & 36.74 $\pm$ 14.32 & 31.77 $\pm$ 14.44 & <0.001 \\
      ODI 3\% (events\,h$^{-1}$)                   & 7.68 $\pm$ 10.32  & 6.30 $\pm$ 9.72   & <0.001 \\
      ODI 4\% (events\,h$^{-1}$)                   & 4.43 $\pm$ 7.33   & 3.67 $\pm$ 7.20   & <0.001 \\
      \bottomrule
    \end{tabular}
  \end{adjustbox}
\end{table}

\subsection{Median-split stratification by nocturnal mean heart rate}
\label{app:joint_tables_hr_median}

\begin{table}[H]
  \centering
  \small
  \setlength{\tabcolsep}{6pt}
  \renewcommand{\arraystretch}{1.15}
  \caption{Full metric comparison between strata defined by the median nocturnal mean heart rate (64\,bpm): high HR (\(\geq\)64\,bpm) vs low HR (\(<\)64\,bpm). Values are mean $\pm$ SD. Overall, differences across most sleep phenotypes were modest under this stratification, suggesting limited discriminative utility of nocturnal mean heart rate alone in the pooled multi-cohort sample. Given the substantial burden of sleep disturbances in the included cohorts, mean heart rate may reflect multiple interacting sleep- and cardiovascular-related factors, thereby reducing the specificity of a single-variable partition. Stratification by nocturnal maximum and minimum heart rate similarly did not yield more distinctive sleep-phenotype separation.}
  \label{tab:joint_hr_median_full}
  \begin{adjustbox}{max width=\linewidth}
    \begin{tabular}{lccc}
      \toprule
      \textbf{Metric} &
      \textbf{High HR (\(\geq\)64) (n=5301)} &
      \textbf{Low HR (\(<\)64) (n=5138)} &
      \textbf{P-value} \\
      \midrule
      Age (years)                 & 64.80 $\pm$ 13.65 & 68.69 $\pm$ 12.14 & <0.001 \\
      BMI (kg\,m$^{-2}$)                 & 28.86 $\pm$ 5.77  & 27.68 $\pm$ 4.77  & <0.001 \\
      AHI (events\,h$^{-1}$)    & 20.01 $\pm$ 18.07 & 18.76 $\pm$ 15.84 & 0.087 \\
      ArI (events\,h$^{-1}$)                       & 29.58 $\pm$ 16.42 & 29.21 $\pm$ 15.01 & 0.695 \\
      N1 \% of TST              & 7.30 $\pm$ 6.56   & 7.25 $\pm$ 6.10   & 0.292 \\
      N2 \% of TST              & 57.79 $\pm$ 11.81 & 58.62 $\pm$ 10.96 & <0.001 \\
      N3 \% of TST              & 15.75 $\pm$ 11.79 & 14.36 $\pm$ 11.06 & <0.001 \\
      NREM \% of TST            & 80.84 $\pm$ 6.74  & 80.22 $\pm$ 6.55  & <0.001 \\
      REM \% of TST             & 19.16 $\pm$ 6.74  & 19.78 $\pm$ 6.55  & <0.001 \\
      SE (\%)                        & 68.33 $\pm$ 14.77 & 67.65 $\pm$ 14.15 & <0.001 \\
      SL (min)                  & 51.47 $\pm$ 55.00 & 54.72 $\pm$ 58.87 & 0.025 \\
      TIB (min)                 & 526.84 $\pm$ 76.27& 538.72 $\pm$ 81.92& <0.001 \\
      TST (min)                 & 354.87 $\pm$ 70.07& 358.38 $\pm$ 65.35& 0.081 \\
      TTSP (min)                & 435.02 $\pm$ 71.77& 443.41 $\pm$ 70.20& <0.001 \\
      WASO (min)                & 80.15 $\pm$ 60.39 & 85.03 $\pm$ 59.68 & <0.001 \\
      Wake \% of TIB            & 31.67 $\pm$ 14.77 & 32.35 $\pm$ 14.15 & <0.001 \\
      ODI 3\% (events\,h$^{-1}$)                   & 6.81 $\pm$ 10.62  & 5.90 $\pm$ 8.74   & <0.001 \\
      ODI 4\% (events\,h$^{-1}$)                   & 4.04 $\pm$ 7.94   & 3.36 $\pm$ 6.34   & <0.001 \\
      \bottomrule
    \end{tabular}
  \end{adjustbox}
\end{table}

\subsection{HRV quantile-based stratification: SDANN as an example}
\label{app:joint_tables_hrv_sdann}

\begin{table}[H]
  \centering
  \small
  \setlength{\tabcolsep}{6pt}
  \renewcommand{\arraystretch}{1.15}
  \caption{Full metric comparison between HRV strata defined by SDANN quantiles: high SDANN (P75; \(\geq\)198.19) vs low SDANN (P25; \(\leq\)114.15). Values are mean $\pm$ SD. HRV candidates considered in this study included SDNN, SDANN, SDANNIndex, RMSSD, pNN50, HRV triangular index (HRV\_tri), LF, HF, LFnu, HFnu, and LF/HF; SDANN is reported here as a representative example. Although some comparisons yield small P-values, between-stratum differences in means and the corresponding variability across most sleep phenotypes are generally modest, suggesting limited discriminative separation when stratifying by a single HRV metric alone in the pooled sample.}
  \label{tab:joint_hrv_sdann_full}
  \begin{adjustbox}{max width=\linewidth}
    \begin{tabular}{lccc}
      \toprule
      \textbf{Metric} &
      \textbf{High SDANN (\(\geq\)198.19) (n=2,609)} &
      \textbf{Low SDANN (\(\leq\)114.15) (n=2,610)} &
      \textbf{P-value} \\
      \midrule
      Age (years)                 & 68.38 $\pm$ 12.35 & 65.17 $\pm$ 13.90 & <0.001 \\
      BMI (kg\,m$^{-2}$)                 & 27.84 $\pm$ 5.11  & 28.90 $\pm$ 5.93  & <0.001 \\
      AHI (events\,h$^{-1}$)    & 19.08 $\pm$ 16.73 & 19.71 $\pm$ 17.55 & 0.394 \\
      ArI (events\,h$^{-1}$)                       & 29.23 $\pm$ 15.55 & 29.44 $\pm$ 16.04 & 0.776 \\
      N1 \% of TST              & 6.87 $\pm$ 5.77   & 7.25 $\pm$ 6.76   & 0.360 \\
      N2 \% of TST              & 58.33 $\pm$ 11.27 & 58.28 $\pm$ 11.92 & 0.972 \\
      N3 \% of TST              & 15.13 $\pm$ 11.52 & 15.45 $\pm$ 11.75 & 0.352 \\
      NREM \% of TST            & 80.32 $\pm$ 6.57  & 80.99 $\pm$ 6.67  & <0.001 \\
      REM \% of TST             & 19.68 $\pm$ 6.57  & 19.01 $\pm$ 6.67  & <0.001 \\
      SE (\%)                        & 66.90 $\pm$ 14.18 & 69.60 $\pm$ 14.95 & <0.001 \\
      SL (min)                  & 52.99 $\pm$ 57.29 & 50.50 $\pm$ 54.52 & 0.172 \\
      TIB (min)                 & 538.68 $\pm$ 82.20& 521.79 $\pm$ 74.35& <0.001 \\
      TST (min)                 & 354.52 $\pm$ 66.65& 357.98 $\pm$ 69.73& 0.002 \\
      TTSP (min)                & 437.33 $\pm$ 72.70& 438.47 $\pm$ 69.53& 0.269 \\
      WASO (min)                & 82.81 $\pm$ 60.14 & 80.49 $\pm$ 59.51 & 0.121 \\
      Wake \% of TIB            & 33.10 $\pm$ 14.18 & 30.40 $\pm$ 14.95 & <0.001 \\
      ODI 3\% (events\,h$^{-1}$)                   & 6.33 $\pm$ 9.40   & 6.60 $\pm$ 10.27  & 0.615 \\
      ODI 4\% (events\,h$^{-1}$)                   & 3.70 $\pm$ 6.94   & 3.88 $\pm$ 7.55   & 0.626 \\
      \bottomrule
    \end{tabular}
  \end{adjustbox}
\end{table}


\clearpage
\section{Pittsburgh Sleep Quality Index (PSQI)}
\label{app:psqi}

{
\noindent\centering
\includegraphics[width=\textwidth,height=0.92\textheight,keepaspectratio]{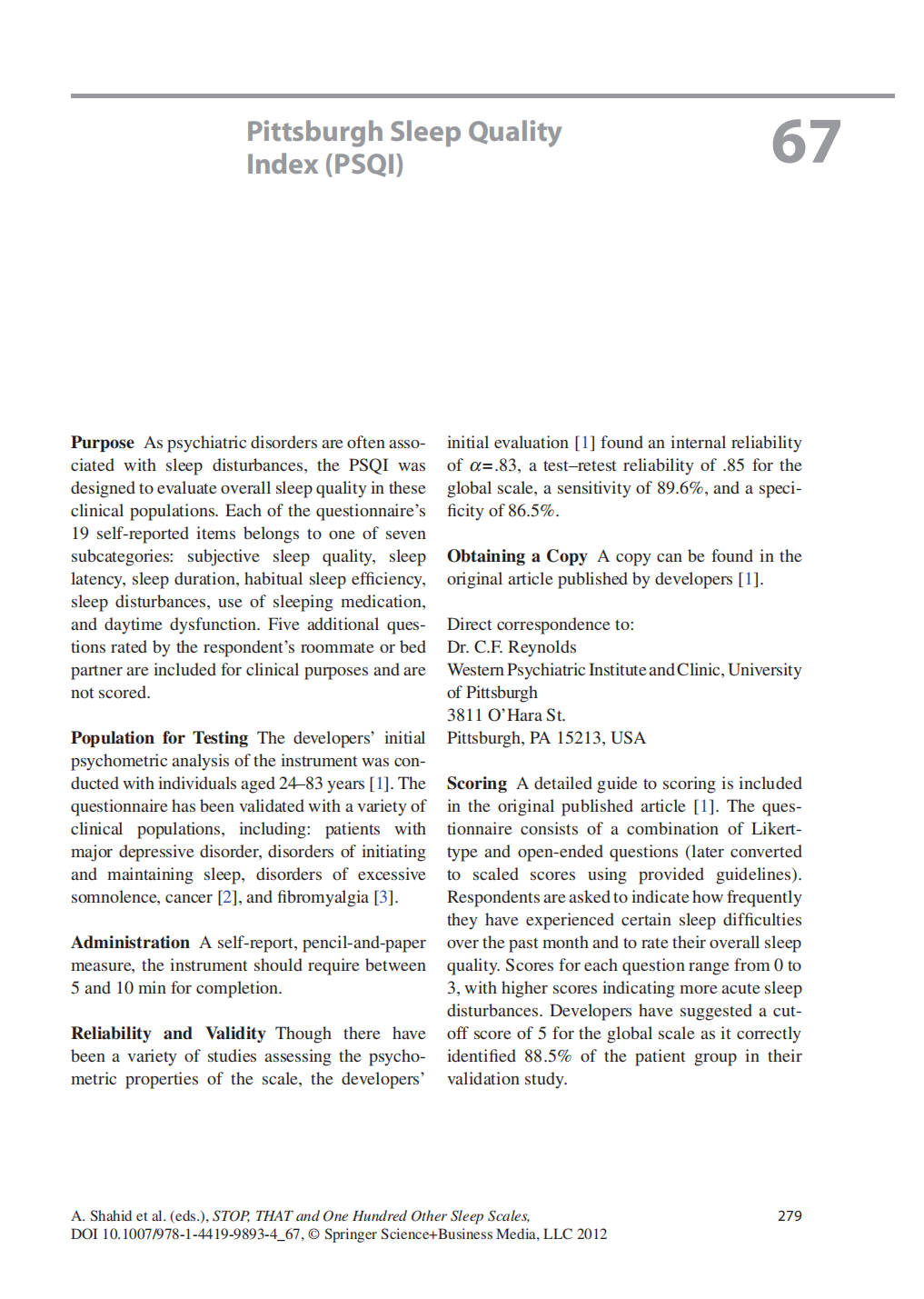}
\par

\clearpage
\noindent\centering
\includegraphics[width=\textwidth,height=0.92\textheight,keepaspectratio]{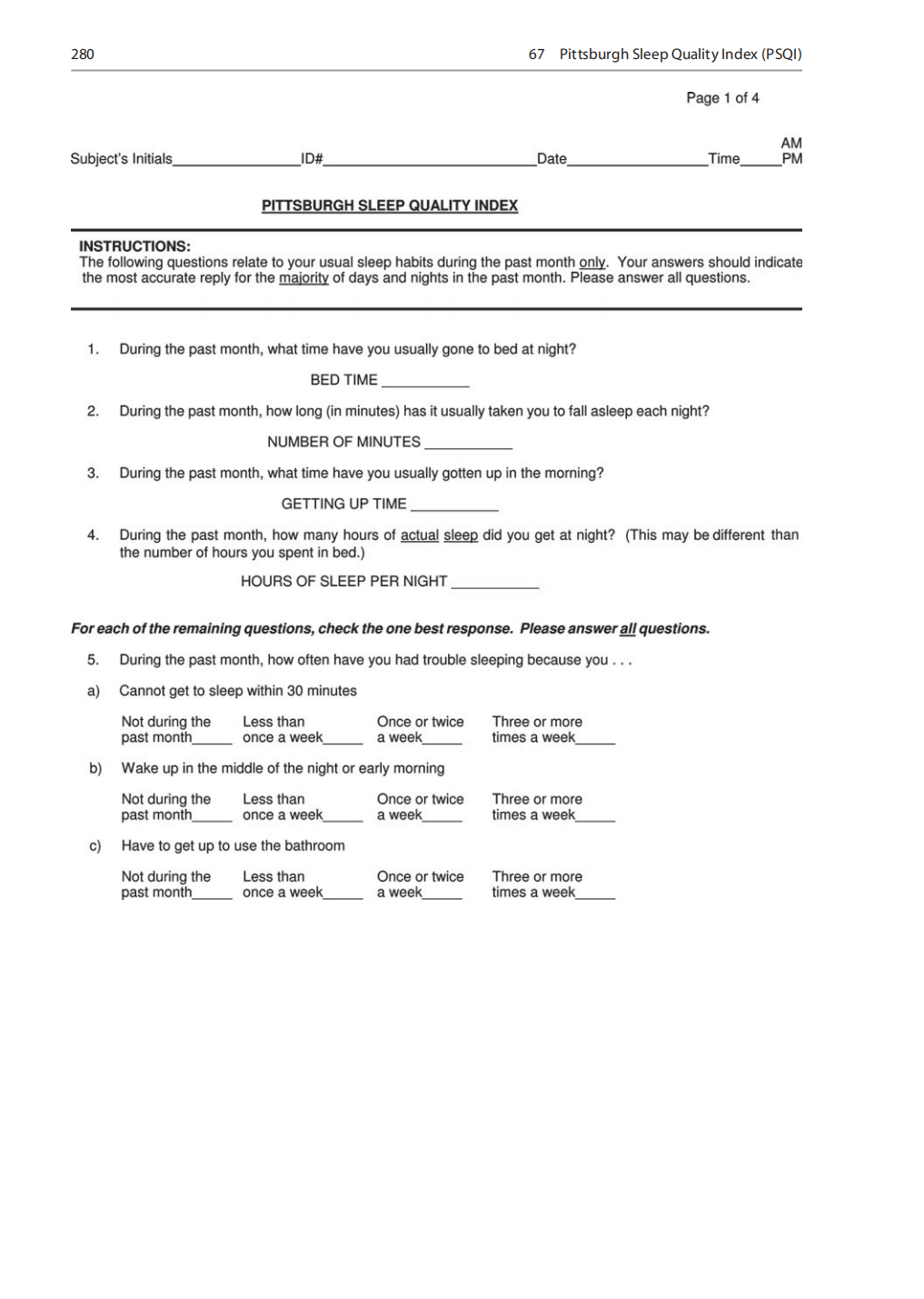}
\par

\clearpage
\noindent\centering
\includegraphics[width=\textwidth,height=0.92\textheight,keepaspectratio]{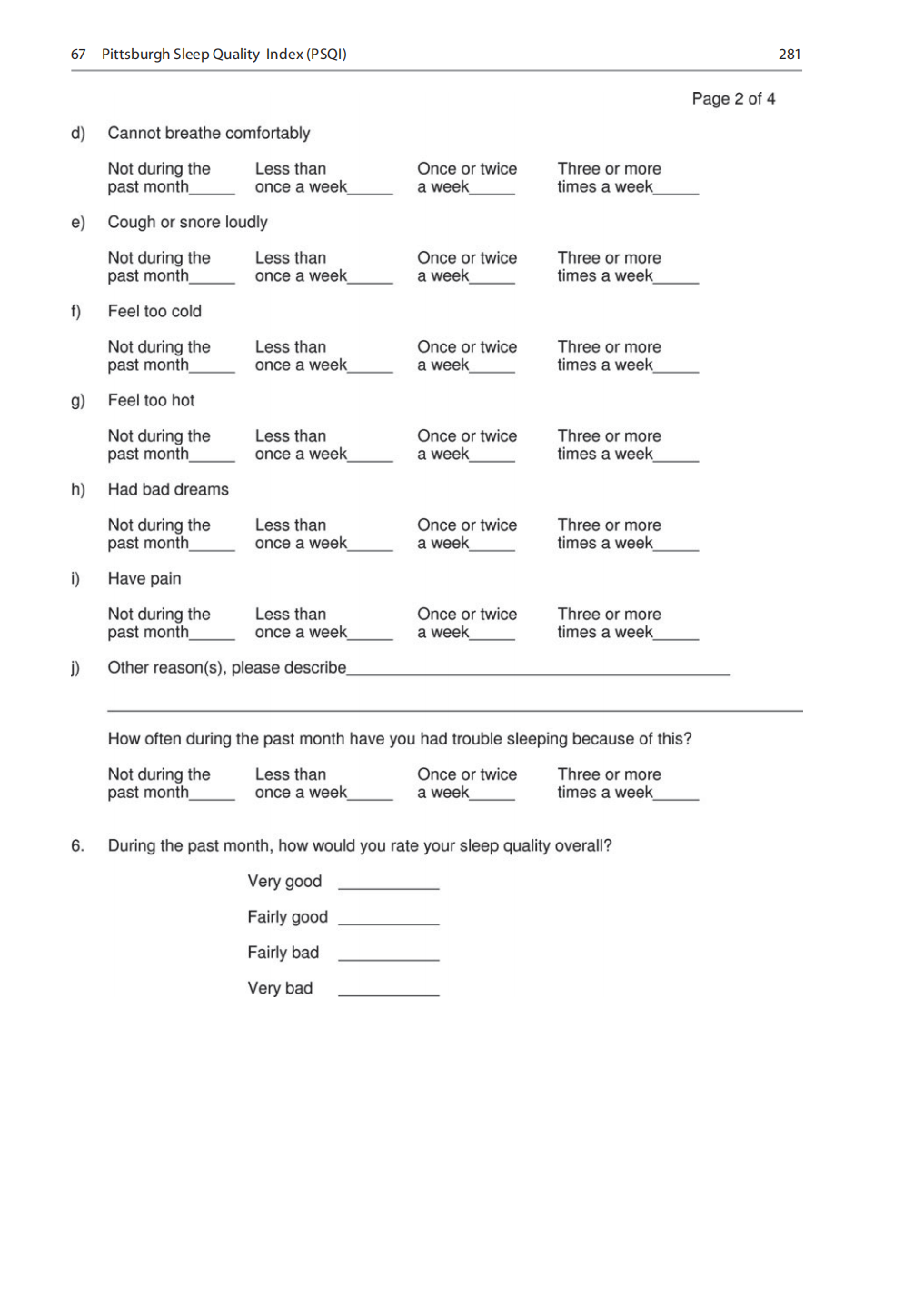}
\par

\clearpage
\noindent\centering
\includegraphics[width=\textwidth,height=0.92\textheight,keepaspectratio]{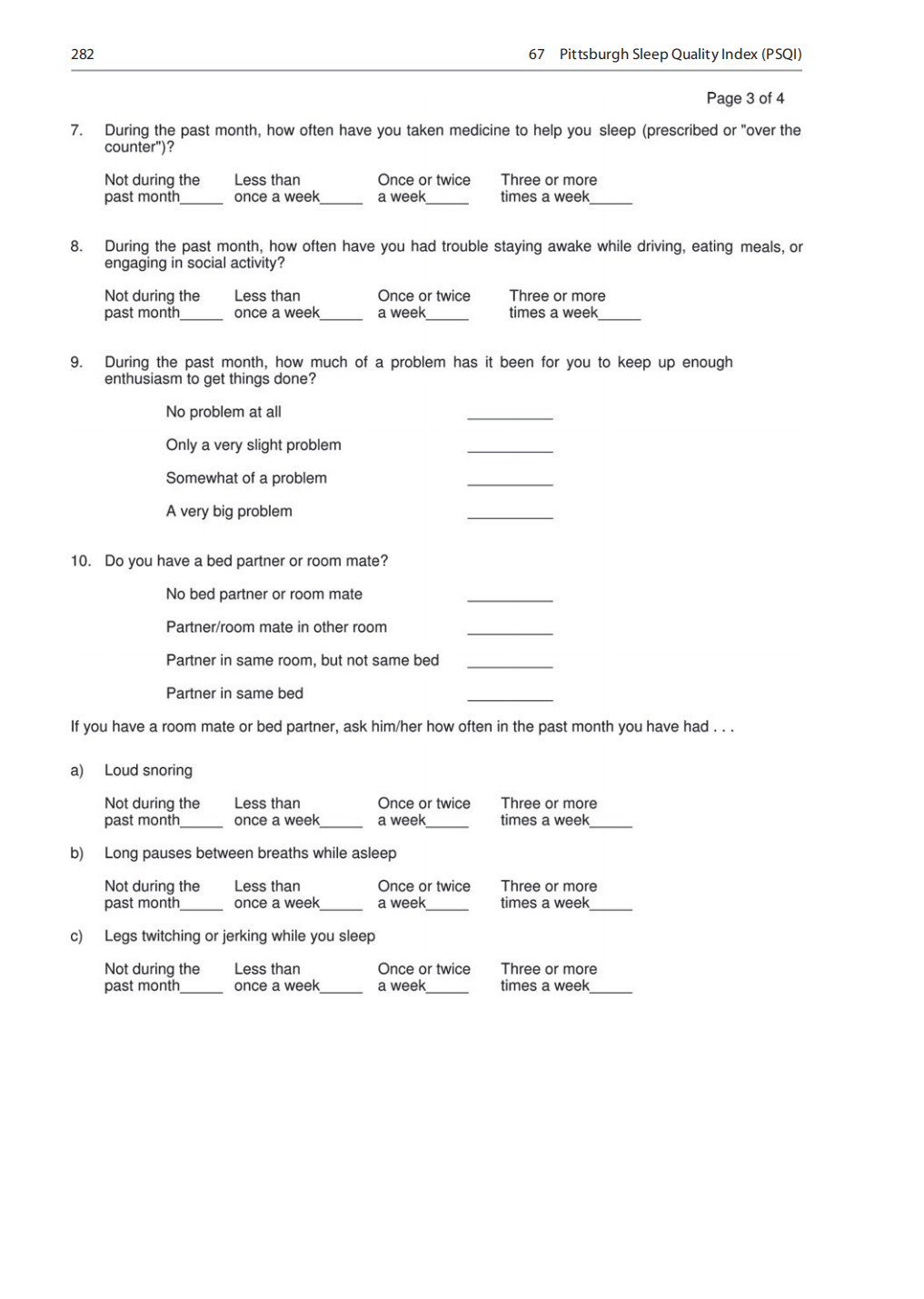}
\par

\clearpage
\noindent\centering
\includegraphics[width=\textwidth,height=0.92\textheight,keepaspectratio]{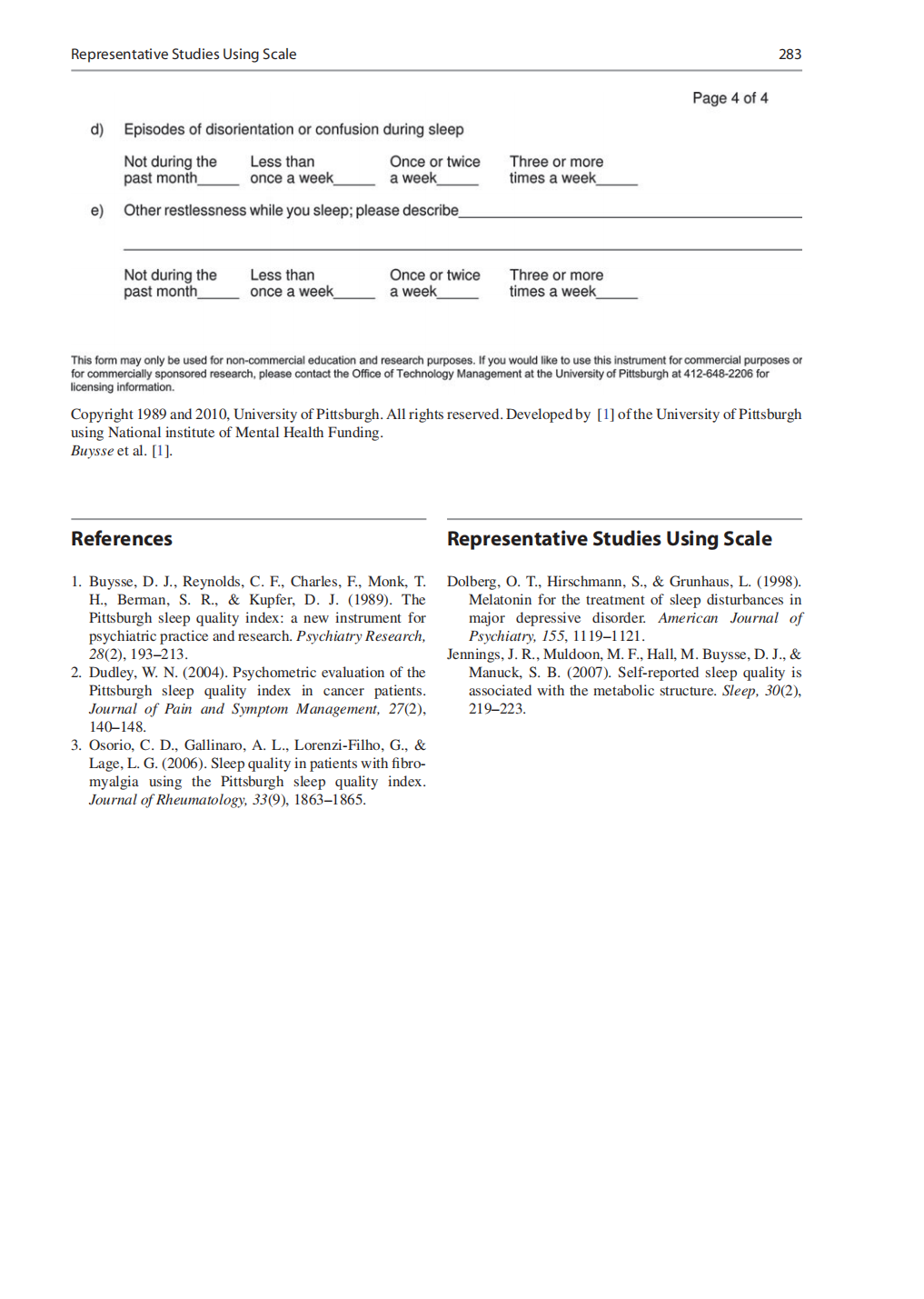}
\par
}

\newpage

\end{appendices}

\end{document}